\documentclass[aps,pra,twocolumn,english]{revtex4}
\usepackage[T1]{fontenc}
\usepackage[latin9]{inputenc}
\usepackage{babel}
\usepackage{graphicx,amsmath,amssymb,color}
\usepackage[normalem]{ulem}
\usepackage{amsfonts}
\usepackage[toc,page]{appendix}
\usepackage{hyperref}
\usepackage{latexsym}
\usepackage{amsfonts}
\usepackage{algpseudocode}
\usepackage{amsthm}
\usepackage{mathrsfs}
\usepackage{natbib}
\usepackage{color,verbatim}
\usepackage{psfrag}
\usepackage{multirow}

\usepackage[svgnames]{xcolor}
\usepackage{tikz}

\usepackage[svgnames]{xcolor}
\usepackage{tikz}
\usetikzlibrary{decorations.markings}
\usetikzlibrary{shapes.geometric}

\pgfdeclarelayer{edgelayer}
\pgfdeclarelayer{nodelayer}
\pgfsetlayers{edgelayer,nodelayer,main}

\tikzstyle{none}=[inner sep=0pt]

\tikzstyle{no_field_qubit}=[fill=white, draw=black, shape=circle,minimum size=0.5cm]
\tikzstyle{plus_field_qubit}=[fill={rgb,255: red,255; green,128; blue,128}, draw=black,shape=circle,label=center:{\tiny \textcolor{black}{$-\lambda$}},minimum size=0.5cm]
\tikzstyle{minus_field_qubit}=[fill={rgb,255: red,128; green,128; blue,255}, draw=black, shape=circle,label=center:{\tiny \textcolor{black}{$+\lambda$}},minimum size=0.5cm]
\tikzstyle{plus_filed_nolabel}=[fill={rgb,255: red,255; green,128; blue,128}, draw=black,shape=circle,minimum size=0.5cm]
\tikzstyle{minus_field_nolabel}=[fill={rgb,255: red,128; green,128; blue,255}, draw=black, shape=circle,minimum size=0.5cm]
\tikzstyle{coloring_node}=[fill=white, draw=black, shape=circle,minimum size=0.5cm]
\tikzstyle{one_hot_node}=[fill={rgb,255: red,255; green,128; blue,255}, draw=black, shape=circle,minimum size=0.3cm]
\tikzstyle{dot_dot_dot}=[fill=none, draw=none, shape=rectangle, label=center:{\Huge \textcolor{black}{...}},minimum size=0.5cm]
\tikzstyle{edge_end}=[fill=none, draw=none, shape=circle]
\tikzstyle{virtual_plus}=[fill={rgb,255: red,255; green,128; blue,128}, draw=black, shape=circle,label=center:{\tiny \textcolor{black}{$-\infty$}},minimum size=0.5cm]
\tikzstyle{virtual_minus}=[fill={rgb,255: red,128; green,128; blue,255}, draw=black, shape=circle,label=center:{\tiny \textcolor{black}{$+\infty$}},minimum size=0.5cm]
\tikzstyle{dw_and_one_hot_qb}=[fill={rgb,255: red,0; green,0; blue,0}, draw=black, shape=circle]
\tikzstyle{one_hot_only_qb}=[fill={rgb,255: red,255; green,0; blue,255}, draw=black, shape=circle]
\tikzstyle{distance_node}=[fill=yellow, draw=black, shape=circle,minimum size=0.5cm]

\tikzstyle{label_1}=[fill=none, draw=none, shape=rectangle]
\tikzstyle{label_2}=[fill=none, draw=none, shape=rectangle]
\tikzstyle{label_3}=[fill=none, draw=none, shape=rectangle]
\tikzstyle{label_4}=[fill=none, draw=none, shape=rectangle]
\tikzstyle{label_5}=[fill=none, draw=none, shape=rectangle]
\tikzstyle{label_6}=[fill=none, draw=none, shape=rectangle]

\tikzstyle{ferro_coupling}=[-,draw=black]
\tikzstyle{anti_ferro_coup}=[-, draw=red]
\tikzstyle{coloring_edge}=[-, draw=black,line width=1]
\tikzstyle{one_hot_edge}=[-, draw={rgb,255: red,0; green,128; blue,0}]
\tikzstyle{domain_wall_edge}=[-,draw=black,line width=2]
\tikzstyle{dw_and_one_hot_edge}=[-, draw={rgb,255: red,0; green,0; blue,0}]
\tikzstyle{one_hot_only_edge}=[-, draw={rgb,255: red,255; green,0; blue,255}]

\newcommand{\ket}[1]{|#1 \rangle}

\newcommand{\average}[1]{\langle #1 \rangle}

\begin{document}

\title{Domain wall encoding of discrete variables for quantum annealing and QAOA}
\author{Nicholas Chancellor}
\email{nicholas.chancellor@gmail.com}
\address{Department of Physics and Durham Newcastle Joint Quantum Centre\\ Durham University, South Road, Durham, UK}

\begin{abstract}

In this paper I propose a new method of encoding discrete variables into Ising model qubits for quantum optimization. The new method is based on the physics of domain walls in one dimensional Ising spin chains. I find that these encodings and the encoding of arbitrary two variable interactions is possible with only two body Ising terms. Following on from similar results for the `one hot' method of encoding discrete variables [Hadfield et. al. Algorithms 12.2 (2019): 34] I also demonstrate that it is possible to construct two body mixer terms which do not leave the logical subspace, an important consideration for optimising using the quantum alternating operator ansatz (QAOA). I additionally discuss how, since the couplings in the domain wall encoding only need to be ferromagnetic and therefore could in principle be much stronger than anti-ferromagnetic couplers, application specific quantum annealers for discrete problems based on this construction may be beneficial. Finally, I compare embedding for synthetic scheduling and colouring problems with the domain wall and one hot encodings on two graphs which are relevant for quantum annealing, the chimera graph and the Pegasus graph. For every case I examine I find a similar or better performance from the domain wall encoding as compared to one hot, but this advantage is highly dependent on the structure of the problem. For encoding some problems, I find an advantage similar to the one found by embedding in a Pegasus graph compared to embedding in a chimera graph.
\end{abstract}

\maketitle


\section{Introduction and background}

There are currently two dominant settings for quantum computing, \emph{gate model} quantum computing, in which computation is realized by a series of discrete `gate' operations, and \emph{continuous time} quantum computing, in which problems are encoded in quantum Hamiltonians and natural dynamics of physical systems are used to find solutions. Furthermore, optimization and statistical sampling have been identified as a potential early application for both gate model and continuous time quantum computing. 

In continuous time quantum computing, optimization is achieved by mapping the optimization problem to the Hamiltonian of a controllable quantum system in such a way that low energy states correspond to more optimal solutions. The most technologically mature continuous time quantum computing devices are the superconducting circuit quantum annealers produced by D-Wave Systems Inc. \cite{D-Wave}. Examples of applications for quantum annealing can be found in diverse fields such as finance\cite{marzec16a,Orus18a,Venturelli18a}, computer science \cite{chancellor16a,choi10b,choi10c}, mathematics\cite{Li17a,Titiloye11a,Titiloye11b}, scheduling \cite{Venturelli15a,crispin13a,Tran16a}, decoding of communications \cite{chancellor16b}, computational biology \cite{perdomo-ortiz12a}, flight gate assignment \cite{Stollenwerk18a}, and air traffic management\cite{Stollenwerk19a}. For gate based machines, one of the most promising algorithms for optimization is the so called the quantum alternating operator ansatz also known as quantum approximate optimization algorithm \cite{Farhi14a,Farhi14b,Yang17a,Jiang17a,Hadfield17a} abbreviated as QAOA. While in principle QAOA could actually be considered in either a gate model or continuous time setting, I restrict the discussion here to gate model implementations.

Like quantum annealing, QAOA requires the optimization problem to be effectively mapped to a Hamiltonian. Gate model quantum computing is less technologically mature, so real world use cases have not been examined \footnote{Technically speaking continuous time formulations of QAOA are possible, but no large scale hardware exists with this capability.} to the extent they have in quantum annealing, although in principle QAOA (or potentially hybrid quantum/classical QAOA based algorithms for thermal sampling) could be applied to many if not all of the applications given previously for quantum annealing. 

A common method of mapping classical optimization problems to quantum hardware is by encoding it into an Ising Hamiltonian,
\begin{equation}
H_{\mathrm{Ising}}=\sum_{i<j}J_{ij}Z_iZ_j+\sum_ih_iZ_i, \label{eq:H_ising}
\end{equation}
where $Z$ is a Pauli $Z$ matrix and $Z_i=\openone_2^{\otimes i-1}\otimes Z \otimes \openone_2^{\otimes n-i}$ where $n$ is the total number of qubits, and $\openone_2$ is the $2\times 2$ identity matrix. In this paper I will focus on encoding into Ising models. Owing to its relative simplicity to experimentally implement, the most common physical implementation of Ising model quantum optimization is the transverse field Ising model
\begin{equation}
H_{\mathrm{trans}}=-A\,\sum_i X_i+B\,H_{\mathrm{Ising}} \label{eq:H_trans}
\end{equation}
where $A$ and $B$ are positive, possibly time dependent, constants and $X_i$ is defined similarly to $Z_i$.

While the challenges in quantum annealing and (gate based) QAOA are not identical, it is likely that both techniques will face some challenges which are similar \cite{Hadfield17a}. It is therefore natural to consider that some of the problem mapping techniques for quantum annealing may be useful in QAOA and vice-versa. In this work I give a technique to efficiently map discrete variables and their interactions to qubits. This method is likely to be useful for both quantum annealing and QAOA, and I will discuss both potential applications.

In the near term, both gate model and quantum annealing devices are likely to have limited connectivity \footnote{One potential exception is optical annealing devices designed in the spirit of so called `coherent Ising machines' \cite{McMahon16a,Inagaki16a}, which may be able to realize full connectivity, however, it is unclear if it is technically feasible to realize such devices at a large scale without classical feedback which interrupts large scale quantum coherence.}. The effects of limited connectivity are different in both cases, for gate model machines, qubit information can be effectively swapped to realize necessary interactions, but this process will increase both circuit depth and gate count, which will be a major concern in near term devices with imperfect gates and limited coherence time. 

For quantum annealing, a more highly connected graph can be realized by minor embedding \cite{choi08a,choi10a} in which logical variables are mapped over strongly interacting `chains' of qubits which form graph minors. Problems can than be mapped to the effective interaction graph of the minors rather than the original graph. Alternatively, the logical variables could be encoded into the parity of qubits \cite{Lechner15a,Rocchetto16a,Albash16b}. For quantum annealing, either embedding technique effectively reduces the number of qubit variables a machine can simulate and decreases the effective dynamic range of energies which can be used in encoding the problem, since both methods require strong interactions to enforce that the qubits remain in logically valid states. Since minor embedding is the most common technique currently used to map problems experimentally, it will be the basis of the analysis in this paper.

There has recently been a significant effort by D-Wave Systems Inc.~to improve the connectivity of their hardware graph to reduce the overheads associated with hardware embedding. The proposed new graph family, known as `Pegasus', is significantly more connected that the current `chimera' graph family \cite{Boothby19a} (for an alternative construction see \cite{Dattani19a}). While hardware with the Pegasus topology is not yet publicly available, Pegasus graphs of various sizes can be generated using the publicly available D-Wave networkx package \cite{D-Wave_nx}. 

Since most quantum hardware is based on quantum bits (qubits), and many optimization problems involve discrete rather than binary variables, one often also needs to encode a higher than binary discrete variable ($\mathbb{Z}_{>2}$) into multiple quantum bits. Discrete variables include integer variables, but can also include other problem representations, including discretized versions of continuum variables, and any case where there are multiple mutually exclusive options. One of many important example of discrete  problem is scheduling, where time can be divided into discrete chunks, and a number of potentially conflicting tasks need to be performed \cite{Venturelli15a,crispin13a,Tran16a}. The time at which each task is performed can be thought of as a discrete variable. Another important example is graph colouring, which seems like a rather esoteric problem but actually has applications in aircraft scheduling, organizing file transfer between processors, and radio frequency assignments \cite{Marx2004a,Titiloye11a,Titiloye11b}. 

Strictly speaking the most informationally dense way to encode such a variable into qubits is to map each value to a binary string, such that a discrete variable of size $m$, (belonging to $\mathbb{Z}_m$ in mathematical language) could be encoded in $\lceil \log_2(m) \rceil$ qubits. Consider the specific case of the interaction between two variables, $V_1\in \mathbb{Z}_m,V_2 \in \mathbb{Z}_m$ using the binary encoding, for simplicity let us restrict ourselves to the case where $m$ is a binary number. In this case, quadratic interactions, including quadratic interactions of a variable with itself ($V_1^2$) will only require second order couplings, since there are $2\,\log_2(m)$ qubits, the means $2\,\log_2(m)\,(2\,\log_2(m)-1)$ couplers. For such interactions, a binary representation is preferable to the encodings discussed here.

However, for higher-than-binary discrete variables, quadratic interactions are a very restrictive form of interactions, general interactions would have to be expressed as a polynomial of order $2\,\log_2(m)=\log_2(m^2)$. Since a polynomial term of order $k$ will require $k$th order couplings, and will in general require these to couple every combination of digits in each binary number. By combinatorics, the total number of such couplers will be $ \sim 2^{\log_2(m^2)}=m^2$. Because building a more than two body interactions out of two body Ising interactions requires at least one auxilliary qubit, the cost of implementing arbitrary interactions between binary representations actually scales worse than less informationally dense encodings. The traditional encoding for arbitrary discrete variable interactions is the \emph{one hot} encoding, which requires $m$ qubits to represent a variable, and does not require additional qubits to represent two variable interactions.

The one hot encoding can be derived by realizing that, if a Hamiltonian is symmetric with respect to exchange of the qubits, than the energy can be written as a function of $\mathbf{Z}=\sum_i Z_i$,  or equivalently, the count of qubits in the $\ket{1}$ configuration, $\mathbf{b}=\sum_i\frac{1}{2}(1-Z_i)$.  The energy with respect to a symmetric Hamiltonian is therefore a polynomial in $\mathbf{b}$, the one hot condition can be enforced by constructing a second order polynomial where the minimum occurs at $\mathbf{b}=1$. In other words, since $\mathbf{b}$ is symmetric, it can be treated as a single parameter which counts the number of variables in the $1$ configuration, a polynomial can be constructed which is uniquely minimized when exactly one variable is in this configuration, this term can effectively act as a constraint,
\begin{equation}
H_{\mathrm{one\,hot},\mathbf{b}}=\lambda \left( \mathbf{b}-1\right)^2=\lambda \left( \mathbf{b}^2-2 \mathbf{b}+1\right),\label{eq:one_hot_b}
\end{equation}
where $\lambda$ is a suitably large positive constant. Translating back into $Z_i$ and dropping irrelevant constant factors yields
\begin{equation}
H_{\mathrm{one\,hot}}=\lambda \left(\sum_{i<j}Z_iZ_j-(m-2)\sum_iZ_i\right),\label{eq:one_hot}
\end{equation}
where $\lambda$ is a suitably large positive constant which enforces that the system should be found in the logically valid subspace with high probability. Because each logical state corresponds to a specific qubit being in the one orientation, arbitrary pairwise interactions between one hot encoded variables can be achieved using two body Ising interactions. 

It is worth briefly commenting that a computer consisting of a single large $\mathbb{Z}_m$ variable is a unary encoding and is therefore not efficient, multiple such variables however have a robust tensor product structure as discussed in \cite{Blume-Kohout02a}, and therefore can be used efficiently in quantum computing, provided there are sufficiently many compared to the (typical) variable size.

I propose an alternative to the one hot encoding based on one dimensional Ising domain walls. I argue that this new encoding is likely to be more useful in near term applications because it requires fewer qubits and in many realistic problem structures also requires a less connected interaction graph, which can lead to more efficient implementations, I also discuss a variety of other advantages. Once I have developed the domain wall encoding, I give a comparison between binary, one hot, and the domain wall encoding in table \ref{tab:comparison}. I discuss both quantum annealing and QAOA implementations, and in particular demonstrate that for several realistic classes of synthetic problems, the domain wall encoding can lead to significant improvement in embedding efficiency over one hot, which in some cases is similar to the comparative advantage of embedding in a Pegasus graph versus a chimera graph.   

The structure of this paper is as follows. In section \ref{sec:dw}, I introduce the domain wall encoding of discrete variables. In the following section, section \ref{sec:dw_int} I introduce how to encode arbitrary two variable interactions similarly to what is known for one hot. In section \ref{sec:mixers}, I discuss how QAOA mixers which do not allow the system to leave the logically valid subspace may be implemented, as has been previously done for one hot in \cite{Hadfield17a}. In section \ref{sec:spec_ann} I discuss advantages of building application specific special purpose annealers which encode discrete problems using the domain wall encoding. In the next section, section \ref{sec:emb_adv} I discuss the advantages of the domain wall encoding for minor embedding in quantum annealing and circuit compilation in QAOA. In section \ref{sec:exp} I provide evidence of the advantage gained from domain wall encoding over one hot for embedding three realistically structured problem types in graphs which are relevant to quantum annealing. In section \ref{sec:methods} I review my numerical methods in the interest of transparency and reproducibility (all code is also publicly available at \cite{domain_wall_code}, including simplified code to implement a domain wall encoding). Finally in section \ref{sec:discuss} I discuss the results and make concluding remarks.

\section{Domain walls in the one dimensional Ising model\label{sec:dw}}

Let us consider the one dimensional ferromagnetic Ising model, defined by the following Hamiltonian
\begin{equation}
H_{\mathrm{inf}}=-\lambda \sum_{i=-\infty}^{\infty} Z_i Z_{i+1}
\end{equation}
where $Z_i$ is a Pauli $Z$ matrix acting on the $i$th qubit.

I define a domain wall as existing between two qubit $i$ and $i+1$ when the bit values of the two qubits are not equal. In other words, doman walls exist for classical basis states where $\langle Z_iZ_{i+1} \rangle =-1$. For a one dimensional ferromagnetic chain of qubits, the energy is simply proportional to the expectation value of the domain wall number. Moreover, a single bit flip on qubit $i$, $X_i$ (as implemented by the driver in Eq.~\ref{eq:H_trans}), can have three different possible effects on domain walls
\begin{enumerate}
\item If $\langle Z_{i-1} \rangle=\langle Z_{i+1} \rangle=\langle Z_i \rangle$, then $X_i$ will create two domain walls, and increase the energy by $4\lambda$
\item If $\langle Z_{i-1} \rangle=\langle Z_{i+1} \rangle\neq \langle Z_i \rangle$, then two domain walls already exist adjacent to qubit $i$ and they annihilate, decreasing the energy by $4\lambda$
\item If $\langle Z_{i-1} \rangle \neq \langle Z_{i+1} \rangle$, than one domain wall exists adjacent to qubit $i$ and a bit flip will move the domain wall at no energy cost
\end{enumerate}
An important consequence of the above is that bit flips cannot remove a single domain wall, a domain wall can only be removed if it encounters another domain wall. In this sense domain walls are \emph{topologically stable}.  In other words, if $\langle Z_{-1} \rangle \neq \langle Z_{N} \rangle$ than there must be an odd number of domain walls between qubits $-1$ and $N$, regardless of the configuration of the intermediate qubits.  This idea is somewhat reminiscent of how (classical) magnetic storage media store classical data in magnetic domains which are topologically protected from local fluctuations as long as they are smaller than the size of the domain \cite{Janzen18a}. Unlike a magnetic hard drive, only a single domain wall per chain is topologically protected, as opposed to one domain wall every time a logical zero is adjacent to a logical one in the hard drive case. The reason for this difference is that in quantum optimisation the fluctuations play an active role in the computation, and therefore simply making a domain large, without logically fixing its value, risks the domain, and therefore the information it contains, being destroyed.

\begin{figure}
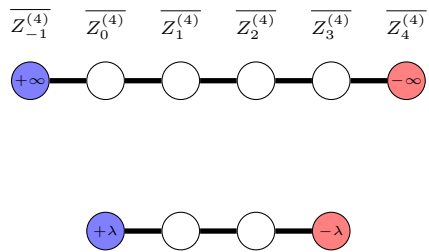

\begin{centering}
\input{tikzit/domain_wall_virtual.label.tex} 
\begin{tikzpicture}
	\begin{pgfonlayer}{nodelayer}
		\node [style={virtual_minus}] (0) at (-4, 3) {};
		\node [style={no_field_qubit}] (1) at (-3, 3) {};
		\node [style={no_field_qubit}] (2) at (-2, 3) {};
		\node [style={no_field_qubit}] (3) at (-1, 3) {};
		\node [style={no_field_qubit}] (4) at (0, 3) {};
		\node [style={virtual_plus}] (5) at (1, 3) {};
		\node [style={minus_field_qubit}] (6) at (-3, 1) {};
		\node [style={no_field_qubit}] (7) at (-2, 1) {};
		\node [style={no_field_qubit}] (8) at (-1, 1) {};
		\node [style={plus_field_qubit}] (9) at (0, 1) {};
		\node [style={label_1}] (10) at (-4, 3.75) {};
		\node [style={label_2}] (11) at (-3, 3.75) {};
		\node [style={label_3}] (12) at (-2, 3.75) {};
		\node [style={label_4}] (13) at (-1, 3.75) {};
		\node [style={label_5}] (14) at (0, 3.75) {};
		\node [style={label_6}] (15) at (1, 3.75) {};
	\end{pgfonlayer}
	\begin{pgfonlayer}{edgelayer}
		\draw [style={domain_wall_edge}] (0) to (1);
		\draw [style={domain_wall_edge}] (1) to (2);
		\draw [style={domain_wall_edge}] (2) to (3);
		\draw [style={domain_wall_edge}] (3) to (4);
		\draw [style={domain_wall_edge}] (4) to (5);
		\draw [style={domain_wall_edge}] (6) to (7);
		\draw [style={domain_wall_edge}] (7) to (8);
		\draw [style={domain_wall_edge}] (8) to (9);
	\end{pgfonlayer}
\end{tikzpicture}
\input{tikzit/blank.label.tex} 
\par
\caption{\label{fig:domain_wall_virtual} Top: One dimensional ferromagnetic Ising chain with end spins held in place by infinitely strong fields. Bottom: Equivalent model where the action of the end spins which cannot change their orientation is substituted by fields. As Table \ref{tab:dw_number} depicts, the degenerate ground state manifold of this model encodes a variable in $\mathbb{Z}_5$.}
\end{centering}
\end{figure}

Let us consider the situation where infinitely strong penalties hold qubit $-1$ in the $1$ state and qubit $N$ in the $0$ state, as depicted in Fig.~\ref{fig:domain_wall_virtual}(top). In this case, there must be an odd number of domain walls between these two fixed qubits. The lowest possible energy in this situation is achieved by placing a single domain wall between any of the $N+1$ pairs of successive qubits, and therefore can be thought of as encoding n discrete variable $x \in \mathbb{Z}_{N+1}\rightarrow \{0,1...N\}$.

We now observe that since qubits $-1$ and $N$ are both effectively fixed by strong constraints, they can be ignored, leaving only a segment of $N$ qubits (indexed between $0$ and $N-1$) with the Hamiltonian
\begin{equation}
H_{\mathrm{N}}=-\lambda [\sum_{i=0}^{N-2} Z_i Z_{i+1}-Z_0+Z_{N-1}],\label{eq:dw_Ham_Z}
\end{equation}
as depicted in Fig.~\ref{fig:domain_wall_virtual}(bottom). The concept of the two `virtual' qubits at the end of the chain allows a useful definition which will simplify the formulation of Hamiltonians later on, I define
\begin{equation}
\overline{Z^{(N)}_i}=
\begin{cases} 
Z_i &  -1<i<N \\
-\openone & i=-1\\
\openone & i=N \\
\mathrm{undefined} & \mathrm{otherwise}
\end{cases}, 
\end{equation}
using this definition, we are able to simplify Eq.~\ref{eq:dw_Ham_Z}
\begin{equation}
H_{\mathrm{N}}=-\lambda \sum_{i=-1}^{N-1} \overline{Z^{(N)}_i}\, \overline{Z^{(N)}_{i+1}}.\label{eq:dw_Ham_Zbar}
\end{equation}
The single domain wall states which are used to encode $\mathbb{Z}_5$ in Fig.~\ref{fig:domain_wall_virtual}(bottom) appear in table \ref{tab:dw_number}. The information storage in the domain wall encoding is a unary encoding (up to the padding with zeros) which, in itself is not a novel way of storing information, although I am not aware of any use of unary encodings in currently used Ising mappings. The novelty of the domain wall encoding comes not from how the information is stored, but how it is processed, in particular, the efficient implementation of two variable interactions discussed in section \ref{sec:dw_int}.

\begin{table}
\begin{centering}
\begin{tabular}{|c|c|}
\hline 
encoded value & qubit configuration\tabularnewline
\hline 
\hline 
0 & 0000\tabularnewline
\hline 
1 & 1000\tabularnewline
\hline 
2 & 1100\tabularnewline
\hline 
3 & 1110\tabularnewline
\hline 
4 & 1111\tabularnewline
\hline 
\end{tabular}
\par
\end{centering}
\caption{\label{tab:dw_number} The five states used to encode $\mathbb{Z}_5$ using the domain wall scheme depicted in Fig.~\ref{fig:domain_wall_virtual}(bottom) }

\end{table}

Additional terms can be used to modify the energy given a domain wall at a given site, this can be accomplished by observing that,
\begin{equation}
\frac{1}{2}\langle (\overline{Z^{(N)}_{i}}-\overline{Z^{(N)}_{i-1})} \rangle=
\begin{cases}
0 &  \langle \overline{Z^{(N)}_{i-1}} \rangle=\langle \overline{Z^{(N)}_{i}} \rangle\\
1 &  \langle \overline{Z^{(N)}_{i-1}} \rangle=-1,\,\,\langle \overline{Z^{(N)}_{i}} \rangle=1\\
-1 &  \langle \overline{Z^{(N)}_{i-1}} \rangle=1,\,\,\langle \overline{Z^{(N)}_{i}} \rangle=-1
\end{cases}, \label{eq:z_bar_def}
\end{equation}
furthermore, it is not possible to have the case where $\langle \overline{Z^{(N)}_{i-1}} \rangle=1,\,\,\langle \overline{Z^{(N)}_{i}} \rangle=-1$ with only a single domain wall, therefore, we define 
\begin{equation}
\bar{\delta}_i=\frac{1}{2} (\overline{Z^{(N)}_{i}}-\overline{Z^{(N)}_{i-1}}) \label{eq:delta_def},
\end{equation}
which assigns an energy penalty of $1$ to domain wall location $i$ and does nothing otherwise (neglecting constant energy offsets, which are discussed later). Using these terms, arbitrary energies can be assigned to any domain wall position. I further observe that a binary variable in the domain wall formalism reduces to a single qubit, and $\overline{Z^{(1)}_0}=Z$, therefore the domain wall formalism recovers the standard binary qubit representation when $N=1$.

It is worth remarking here that we could instead define $\bar{\delta}_i$ in an arguably more natural way using two body operations
\begin{equation}
\bar{\delta}'_i=\frac{1}{2} (1-\overline{Z^{(N)}_{i}}\,\overline{Z^{(N)}_{i-1}}) \label{eq:delta_alt_def},
\end{equation}
however, while this definition is completely mathematically valid, I will show later that this requires four body Ising interactions to implement arbitrary two body interactions between the encoded variables.

\section{Interactions between domain wall variables \label{sec:dw_int}}

Now that I have demonstrated how to asign penalties for single discrete variables, I move on to discuss coupling between domain wall encoded variables. To do this, I must first introduce notation for additional variables, this is accomplished by introducing a second index relating to the variable number, $k$, $\bar{\delta}_{i}\rightarrow \bar{\delta}_{i}^k $ and $\overline{Z^{(N)}_i} \rightarrow \overline{Z^{(N),k}_i}$.

I now observe that the term $\bar{\delta}_{i}^k \, \bar{\delta}_{j}^l $ is one iff variable $k$ has a value of $i$ and variable $l$ has a value of $j$, and is zero for all other values of variables $k$ and $l$. From this observation, it follows that an arbitrary two variable function can be created from\footnote{Again up to a constant energy offset, see later discussion. Note that the single variable terms can be created from sums of two variable terms.}
\begin{equation}
H_\mathrm{2 var}=\sum_{i=0}^N\sum_{j=0}^M E_{i,j}\bar{\delta}_{i}^k \, \bar{\delta}_{j}^l  \label{eq:2var}
\end{equation}
Furthermore, by substituting in Eq.~\ref{eq:delta_def} the product
\begin{align}
\bar{\delta}_{i}^k \, \bar{\delta}_{j}^l =\frac{1}{4}(\overline{Z^{(N),k}_{i-1}}\,\overline{Z^{(M),l}_{j-1}} \nonumber \\
-\overline{Z^{(N),k}_{i}}\,\overline{Z^{(M),l}_{j-1}}-\overline{Z^{(N),k}_{i-1}}\,\overline{Z^{(M),l}_{j}}+\overline{Z^{(N),k}_{i}}\,\overline{Z^{(M),l}_{j}}), \label{eq:int_def}
\end{align}
since every term of this equation is at most two-body by Eq.~\ref{eq:z_bar_def}, it immediately follows that arbitrary two variable functions can be constructed by two body Ising couplers between encoded domain wall variables. Furthermore since there are only $N\times M$ possible couplers between an encoded $\mathbb{Z}_{N+1}$ and a $\mathbb{Z}_{M+1}$, it follows that this can be accomplished with at most $N\times M$ two body Ising terms. Since binary variables can be considered a special case of the domain wall encoding, arbitrary coupling between standard binary Ising variables (i.e.~in mixed binary/integer problems) and domain wall encoded discrete variables is possible without requiring any special modification to the formalism. 

Before moving on to applications of  the domain wall encoding, it is important to make one technical mathematical note about the domain wall encoding in contrast to the one hot encoding. If we are using the domain wall encoding to encode an interaction between a $\mathbb{Z}_n$ and a $\mathbb{Z}_m$ variable, then the number of independent Hamiltonian terms used to encode the interaction will be $(m-1)\times(n-1)$ two body interactions and $n+m-2$ single body terms, leaving a total of $n\times m-1$ total independent degrees of freedom to control $n\times m$ independent interaction terms. The missing degree of freedom is accounted for by the fact that all physical dynamics (and the ordering and gaps between energies of solutions) are invariant under a shift in the defined zero of energy. 

A redefinition of the zero of energy provides an additional degree of freedom which is purely mathematical. Non-trivial physical interactions which shift the zero of energy are possible in one hot, a fully connected interaction between all of the qubits in each variable will penalize all $n\times m$ states equally. If one attempts to construct a similar `gauge operator' in the domain wall encoding by summing all possible terms in Eq.~\ref{eq:2var}, all interaction terms from the individual expansion in Eq.~\ref{eq:int_def} will cancel. 

The fact that the number available one and two body interactions plus redefinition of the zero energy exactly equals $n\times m$ implies that the domain wall encoding is the densest possible encoding of arbitrary two variable interactions between integers using only one and two body Ising terms and no auxilliary qubits, using any fewer number of qubits would not leave enough degrees of freedom to arbitrarily control the interaction.

Let us now briefly consider what would happen if I instead had defined interactions using $\bar{\delta}'_i$ from  Eq.~\ref{eq:delta_alt_def}, in that case, the resulting product would be 
\begin{align}
\bar{\delta}_{i}'^k \, \bar{\delta}_{j}'^l =\frac{1}{4}(1-\overline{Z^{(N),k}_{i}}\,\overline{Z^{(N),k}_{i-1}} \nonumber \\
-\overline{Z^{(N),l}_{j}}\,\overline{Z^{(N),l}_{j-1}}+\overline{Z^{(N),k}_{i}}\,\overline{Z^{(N),k}_{i-1}}\,\overline{Z^{(N),l}_{j}}\,\overline{Z^{(N),l}_{j-1}}), \label{eq:int_alt_def}
\end{align}
which requires a four body coupler and is therefore much less convenient to implement. One of the major results in this paper is that the coupling between domain wall variables can be implemented \emph{only using two body coupling} if built from the definition given in Eq.
~\ref{eq:delta_def} rather than the one in Eq.~\ref{eq:delta_alt_def}. For the remainder of this work, I will only consider the interaction encoding defined in Eq.~\ref{eq:int_def} because of the clear advantage it has in only requiring two body couplers to implement.

\section{QAOA mixers\label{sec:mixers}}
Traditionally quantum annealing and QAOA use transverse field mixers described by Hamiltonians of the form  Eq.~\ref{eq:H_trans}. However, this allows for the possibility of ending the call to the protocol in an invalid state, which in the case of the domain wall encoding would be any state where a variable encoding has more than one domain wall. Such states are problematic because they do not uniquely correspond to solutions to the original problem, and while it is possible that an advantage could be obtained using clever post-processing, an invalid state is still an undesirable outcome. The problem of finding invalid states in finite temperature quantum annealing has been highlighted in \cite{Albash17a}. It would therefore be preferable to use a mixing Hamiltonian which only mixes between valid states, as discussed in \cite{Hadfield17a,Marsh18a,Wang19a}. These papers have focused on QAOA, since currently existing quantum annealers use transverse field mixers. There has however been substantial progress \cite{Deng18a,Ozfidan19a} recently on two body mixing terms for quantum annealing, therefore, it may not be outside of the realm of possibility (although probably further in the future) that the mixers proposed in this section could be implemented in quantum annealers.

Recall that flipping a qubit which is adjacent to a single domain wall does not change the domain wall number, therefore, we should construct Hamiltonian terms which only perform a bit flip operation if the qubit is adjacent to a single domain wall \footnote{Strictly speaking we only want to prevent bit flips in the no domain wall case, since the two domain wall case is already an invalid state.}. Fortunately the Hamiltonian term $(Z_{i-1}X_i-X_iZ_{i+1})$ satisfies exactly this property. Starting from any computational basis state where $\langle Z_{i-1} \rangle= \langle Z_{i+1} \rangle$, the two Pauli $X$ terms on qubit $i$ will cancel. On the other hand, for $\langle Z_{i-1} \rangle=-\langle Z_{i+1} \rangle$, the action will be $\pm X$, depending on which side of the qubit the domain wall is located. Summing together such terms for each domain wall site yields
\begin{equation}
H_{\mathrm{mix}}=\sum_{i=0}^{N-1} (\overline{Z^{(N)}_{i-1}}X_i-X_i\overline{Z^{(N)}_{i+1}}). \label{eq:mix_ham}
\end{equation}
While this mixer Hamiltonian contains sums of non-commuting terms, it can be broken down into the sum of two Hamiltonians constructed out of commuting terms. This division works by observing that $X$ and $Z$ terms are always consecutive, therefore Hamiltonian terms with all of their $Z$ components on odd (even) qubits will have $X$ components on even (odd) terms. The Hamiltonian can be split as follows $H_{\mathrm{mix}}=H^\mathrm{even}_{\mathrm{mix}}+H^\mathrm{odd}_{\mathrm{mix}}$ where 
\begin{equation}
H^\mathrm{even}_{\mathrm{mix}}=\sum_{i=0}^{\lfloor \frac{N-1}{2} \rfloor} (\overline{Z^{(N)}_{2\,i-1}}X_{2\,i}-X_{2\,i}\overline{Z^{(N)}_{2\,i+1}}),
\end{equation}
and
\begin{equation}
H^\mathrm{odd}_{\mathrm{mix}}=\sum_{i=0}^{\lceil \frac{N+1}{2} \rceil} (\overline{Z^{(N)}_{2\,i}}X_{2\,i+1}-X_{2\,i+1}\overline{Z^{(N)}_{2\,i+2}}).
\end{equation}
Not only does $H_{\mathrm{mix}}$ conserve domain wall number, but each $H^\mathrm{even\,(odd)}_{\mathrm{mix}}$ both do as well, implying that any operator formed by performing the unitaries created from these Hamiltonians will also conserve domain wall number. Indeed, each of these terms can be constructed from $Z_iZ_{i+1}$ and Hadamards. Since (exponentiated) two body Ising terms are already necessary to produce the phase separators, then this mixer can be efficiently constructed from two body terms which already exist.

While the mixer in Eq.~ \ref{eq:mix_ham} is similar to the controlled-$X$-rotation  mixer discussed at the beginning of 4.2.2 of \cite{Hadfield17a}, there is an important distinction, while the controlled-$X$-rotation mixer is controlled by a single qubit value, whether or not the $X$ is applied in Eq.~ \ref{eq:mix_ham} is actually applied by whether two qubits agree or differ. The approach of splitting the driver into commuting parts which follows after that equation is essentially the same as what was done in \cite{Hadfield17a}.

It is worth observing that the mixers here explore the solution space in a fundamentally different way than the one hot mixers in \cite{Hadfield17a}. Those mixers allow a transition from any state to any other state, whereas the methods proposed here only allow transitions between consecutive states. It is not immediately obvious which of these mixers will actually perform better in real problems. On one hand, it is known that a speedup is not possible for quantum search in too low of a dimension \cite{childs03a}, however, low dimensionality is only problematic in dimensions less than $4$, meaning that if this result carries over from search to optimization (it is not \emph{a priori} clear that it would), then a quantum advantage would be possible in any case with more than $4$ variables, which should be the case in all interesting optimization problems. On the other hand, it has been shown that it problematic to have a mixer which is fully connected, as would be the case for a single one hot variable \cite{farhi08a,chancellor18a,callison19a}. However, since there will be many variables in a real problem, the total mixer graph formed in the solution space in the one hot encoding is likely to be quite far from fully connected. Therefore, while the behaviour of these two mixers is different, which is better for computation should be treated as an open, likely problem dependent, question. 

\section{Specialized discrete optimization annealers \label{sec:spec_ann}}
In addition to considering specialized QAOA mixer Hamiltonians for domain wall encodings, it is also worth briefly discussing the possibility of constructing specialized quantum annealers which are specifically designed for discrete, rather than binary problems. The Hamiltonian to domain wall encode a variable can be constructed entirely from single body `field' terms and ferromagnetic (negative) coupling. Because of the way in which currently used flux qubit couplers are constructed, these terms can be implemented much more strongly than anti-ferromagnetic coupling \cite{van_den_Brink05a}. Therefore, especially for relatively small discrete sets, it may be possible to construct a specialized quantum annealer designed to handle discrete variables with little or no sacrifice in the dynamic range available for problem setting as compared to a binary machine. To some extent, the controls of D-Wave hardware already allow users to take advantage of the ability of ferromagnetic coupling to be stronger, but under the context of minor embedding \cite{virtual_graphs_whitepaper}.

If a transverse field mixer is used, then the system would necessarily access invalid higher energy states, and the energy separation would have to be sufficient that these states are not accessed. In a flux qubit quantum annealer, the Ising spins are already formed from the two lowest energy states of an infinite ladder for each qubit, and modelling these additional states is sometimes important to fully understand the dynamics \cite{johnson11a}. Creating discrete variables with domain wall encodings would therefore not represent a fundamental change to how these devices work. A more exotic and ambitious option would be to develop an annealer which has a mixer Hamiltonian of the form in Eq.~\ref{eq:mix_ham}, however, recall that this would require a significant advance in available mixer terms. Specialized drivers for quantum annealing which act over a feasible subspace has been examined previously for other problem encodings \cite{Hadfield17a,Hen16a,Hen16b}.

Specialized annealers designed to handle discrete problems could be particularly useful if a high value set of problems with similar or identical structure were identified. This would allow for the possibility of an application specific integrated circuit (ASIC) annealer designed to solve specific high value problems. Such an ASIC approach would make it possible to reduce or eliminate the overhead associated with embedding for a family of high value problems, since embedding overhead can greatly reduce performance on current, non-specialized quantum annealers, reducing or eliminating this source of overhead is likely to result in a major increase in performance.

One final advantage of the domain wall encoding as compared to one hot is that the coupling between logical states using a transverse field driver is non-perturbative in the sense that the system does not have to pass through a logically invalid state to get to different logically valid states. The effective transition rates between logical states is therefore independent of the coupling. In contrast, to pass between two logically valid one hot states under transverse field driving, the system must pass through a state with either more than one qubit in the one configuration, or zero qubits in the one configuration. For a fixed transverse field the effective coupling between logically valid states will therefore decrease as the strength of the penalties enforcing the one hot constraint are increased. Not having a tradeoff between encoding strength and coupling strength is likely to make design of specialized domain wall encoded hardware simpler.

\section{Embedding/compilation\label{sec:emb_adv}}
There are several important differences when considering the domain wall encoding proposed here when compared to one hot encoding with respect to minor embedding in the case of quantum annealing, or circuit compilation in the case of QAOA. These differences all relate to the interaction graph structure of the qubits encoding the problem.

The most obvious in terms of embedding overhead is that a domain wall encoding requires one fewer qubit per discrete variable, while nominally a minor improvement, this could be significant when encoding small discrete variables, for instance in a problem composed of $\mathbb{Z}_3$ variables, the qubit count would be reduced by $\frac{1}{3}$. There are, however, more subtle advantages which are likely to be more important. The domain wall encoding requires significantly less connectivity within qubits encoding a variable than one hot. In one hot all of the qubits encoding a variable need to be interconnected, while the domain wall encoding only requires linear connectivity. Finally,  the interactions between the variables will be different in both cases.  In summary, the three differences between the interaction graphs of the two encodings are
\begin{enumerate}
\item The domain wall encoding requires one fewer qubit per discrete variable
\item The domain wall encoding requires only linear connectivity for the qubits used to encode a single discrete variable, while one hot requires full connectivity
\item While both methods can implement arbitrary two variable functions using two body interactions between the qubits encoding the two variables, encoding a \emph{particular} interaction will require different interactions between the qubits, in some cases one hot will require more inter-variable interactions, in others the domain wall encoding will, the interaction structure will also be different
\end{enumerate}

To mathematically capture some of the structural difference between these two strategies, I consider the \emph{edge distance} $d_e$ between qubit variables in a graph, defined simply as the minimum number of edges which must be traversed to get from one vertex to a different vertex. In the one hot encoding, the edge distance between qubits which encode the same variable $\mathbb{Z}_n$ is always $1$, whereas for a domain wall encoding, the edge distance between two such qubits can be as high as $n-1$, depending on other interactions.

Hardware graph connectivity has proven to be a major obstacle in quantum annealing. Recall that the conventional strategy when a problem graph is not a subgraph of a given hardware is to minor embed \cite{choi08a,choi10a} variables by encoding each of them to strongly coupled qubits which form a graph minor. Minor embedding effectively reduces the number of available qubits, and can lead to issues such as `broken' variables due to thermal fluctuations \cite{Albash17a}. For quasi-planar geometries, like the D-Wave chimera graph, the size of fully connected graph which can be represented on a given device goes as the square root of the number of qubits. Parity based encodings \cite{Lechner15a,Rocchetto16a,Albash16b} provide a potential alternative to minor embeddings, but the size of fully connected graph which can be represented in a quasi-planar geometry still scales as the square root of the number of qubits \footnote{By graph theoretical arguments relating to the treewidth, this will actually be true for any encoding strategy}. 

For the readers convenience, I have constructed table \ref{tab:comparison} which lists key performance metrics for binary, one-hot, and domain wall strategies.

\begin{table*}[t]
\begin{tabular}{|c|c|c|c|}
\hline 
performance metric & binary & one hot & domain wall \tabularnewline
\hline 
\hline 
\# qubits & \textcolor{blue}{$\lceil\log_{2}(m) \rceil$} & $m$ & $m-1$ \tabularnewline
\hline 
\# couplers & \textcolor{blue}{$0$ if $m=2^{n}\:n\in\mathbb{Z}$} & \multirow{2}{*}{$m\,(m-1)$} & \multirow{2}{*}{$m-2$} \tabularnewline
for encoding & \textcolor{red}{complicated otherwise} &  &  \tabularnewline
\hline 
intra-variable connectivity & \textcolor{blue}{N/A} or \textcolor{red}{complicated} & \textcolor{red}{complete} & \textcolor{blue}{linear}\tabularnewline
\hline 
maximum order & \multirow{2}{*}{\textcolor{red}{$\,\lceil\log_{2}(m) \rceil$}} & \multirow{2}{*}{\textcolor{blue}{$1$}} & \multirow{2}{*}{\textcolor{blue}{$1$}}\tabularnewline
needed to penalize single values &  &  &  \tabularnewline
\hline 
maximum order & \multirow{2}{*}{\textcolor{red}{$2\,\lceil\log_{2}(m) \rceil$}} & \multirow{2}{*}{\textcolor{blue}{$2$}} & \multirow{2}{*}{\textcolor{blue}{$2$}}\tabularnewline
needed for two variable interactions &  &  &  \tabularnewline
\hline 
maximum $d_e$ between & \multirow{2}{*}{complicated} & \multirow{2}{*}{ \textcolor{red}{$2$}} & \multirow{2}{*}{\textcolor{blue}{$m$}} \tabularnewline
qubits in interacting variables &  &  &  \tabularnewline
\hline 
\end{tabular}

\caption{\label{tab:comparison}Comparison between binary, one hot and  domain wall encoding strategies (note that the $\delta'_{i}$ strategy is not shown in the table, but would be the same as the one used here except for would require fourth order coupling for a two variable interaction). Maximum order in this case refers to the maximum number of $Z$ variables which must appear in a single Hamiltonian term for the encoding. \textcolor{red}{Red} colouring is used to indicate a major drawback of a strategy, while \textcolor{blue}{blue} indicates a major advantage conferred by a strategy. The word `complicated' is used to indicate cases where the result is likely to be highly dependent on the details of the problem being encoded. For discussion of the performance metrics, and explanations of the `complicated' cases, see appendix 1.}

\end{table*}

Gate model quantum machines are less technologically mature, and therefore real world problem embedding strategies (which can be considered part of the circuit compilation problem) are less developed. However, the connectivity of the interaction graph on these devices is also likely to lead to overhead. One strategy to encode interactions which are not natively present in the hardware graph is to perform {\sc swap} operations between neighbouring physical qubits and thereby shuttle logical variables around to achieve necessary interactions (see for example \cite{roffe17a,Whitney07a,Cowtan19a}). These {\sc swap} operations contribute to the total circuit depth of a QAOA implementation. In a fully fault tolerant setting, this would only have the relatively minor consequence of an increased runtime. Near term devices, however, are likely to be far from fault tolerant, and therefore only be able to reliably implement relatively shallow circuits, it is therefore highly desirable to reduce circuit depth. Because the eventual structure of large scale gate based quantum devices is still unclear, I restrict the study of specific examples to embedding in quantum annealing.

\section{Examples \label{sec:exp}}
In this section, I compare domain wall and one hot minor embeddings for three realistic families of problem structures. The goal here is not to generate provably hard problems, but rather to reproduce realistic structures which may be encountered in the real world. In all cases I examine, I find that domain wall encoding yields at least a small advantage over one hot, but that the size of the advantage is highly problem structure dependant.

As part of the study here, I numerically examine embedding into both the D-Wave chimera graph, and the recently proposed Pegasus graph. I find that in the case of synthetic scheduling problems, the advantage in embedding efficiency between one hot and domain wall encodings is comparable to that of embedding into a chimera versus a Pegasus graph. On the other extreme, I find that embedding domain wall encoded maximum three colour problems is actually slightly less efficient than for one hot, but that the requirement of one fewer qubit per variable more than makes up for this difference and still leaves domain wall encoding as the preferred strategy. I first describe the basic results for the three families of problems, before a more in-depth comparative analysis in subsection \ref{sub:analysis}. For transparency and reproducibility, the Hamiltonians for each example are provided in appendix 2.

\subsection{Unstructured Interactions}

Let us consider unstructured interactions, by which I mean interactions for which there is no particular structure which makes the variables independent from each other in certain regimes and therefore require all two body terms to construct the interactions in both the one hot and domain wall encoding \footnote{Strictly speaking, one edge can always be removed by a judicious choice of the zero of energy in one hot, but not in the domain wall encoding, consequences will be discussed in a later footnote.}. Unstructured interactions may come about for example if the interactions between the discrete variables are describing complex correlations, for instance in an discrete analogy to \cite{marzec16a}, but where each discrete variable represents more than two mutually exclusive possibilities.

\begin{figure}
\begin{centering}
\begin{tikzpicture}
	\begin{pgfonlayer}{nodelayer}
		\node [style={dw_and_one_hot_qb}] (1) at (-4.25, 2.25) {};
		\node [style={dw_and_one_hot_qb}] (2) at (-3.25, -0.5) {};
		\node [style={dw_and_one_hot_qb}] (3) at (-3.25, 0.5) {};
		\node [style={one_hot_only_qb}] (10) at (-4.75, 1) {};
		\node [style={dw_and_one_hot_qb}] (11) at (0, 4.5) {};
		\node [style={dw_and_one_hot_qb}] (12) at (-0.5, 3.25) {};
		\node [style={dw_and_one_hot_qb}] (13) at (1.5, 3.25) {};
		\node [style={one_hot_only_qb}] (14) at (-1.25, 4.5) {};
		\node [style={dw_and_one_hot_qb}] (15) at (2, 1.5) {};
		\node [style={dw_and_one_hot_qb}] (16) at (0.75, -0.5) {};
		\node [style={dw_and_one_hot_qb}] (17) at (2.75, 0.25) {};
		\node [style={one_hot_only_qb}] (18) at (-0.25, -1.25) {};
	\end{pgfonlayer}
	\begin{pgfonlayer}{edgelayer}
		\draw [style={domain_wall_edge}] (1) to (2);
		\draw [style={domain_wall_edge}] (2) to (3);
		\draw [style={one_hot_only_edge}] (10) to (1);
		\draw [style={one_hot_only_edge}] (10) to (3);
		\draw [style={one_hot_only_edge}] (10) to (2);
		\draw [style={one_hot_only_edge}] (1) to (3);
		\draw [style={domain_wall_edge}] (11) to (12);
		\draw [style={domain_wall_edge}] (12) to (13);
		\draw [style={one_hot_only_edge}] (14) to (11);
		\draw [style={one_hot_only_edge}] (14) to (13);
		\draw [style={one_hot_only_edge}] (14) to (12);
		\draw [style={one_hot_only_edge}] (11) to (13);
		\draw [style={dw_and_one_hot_edge}] (1) to (11);
		\draw [style={dw_and_one_hot_edge}] (3) to (11);
		\draw [style={dw_and_one_hot_edge}] (2) to (11);
		\draw [style={dw_and_one_hot_edge}] (1) to (12);
		\draw [style={dw_and_one_hot_edge}] (2) to (12);
		\draw [style={dw_and_one_hot_edge}] (3) to (12);
		\draw [style={dw_and_one_hot_edge}] (3) to (13);
		\draw [style={dw_and_one_hot_edge}] (1) to (13);
		\draw [style={dw_and_one_hot_edge}] (2) to (13);
		\draw [style={one_hot_only_edge}] (10) to (14);
		\draw [style={one_hot_only_edge}] (10) to (11);
		\draw [style={one_hot_only_edge}] (10) to (12);
		\draw [style={one_hot_only_edge}] (10) to (13);
		\draw [style={one_hot_only_edge}] (14) to (1);
		\draw [style={one_hot_only_edge}] (14) to (3);
		\draw [style={one_hot_only_edge}] (14) to (2);
		\draw [style={domain_wall_edge}] (15) to (16);
		\draw [style={domain_wall_edge}] (16) to (17);
		\draw [style={one_hot_only_edge}] (18) to (15);
		\draw [style={one_hot_only_edge}] (18) to (17);
		\draw [style={one_hot_only_edge}] (18) to (16);
		\draw [style={one_hot_only_edge}] (15) to (17);
		\draw [style={dw_and_one_hot_edge}] (15) to (13);
		\draw [style={dw_and_one_hot_edge}] (15) to (12);
		\draw [style={dw_and_one_hot_edge}] (15) to (11);
		\draw [style={dw_and_one_hot_edge}] (17) to (1);
		\draw [style={dw_and_one_hot_edge}] (17) to (3);
		\draw [style={dw_and_one_hot_edge}] (17) to (2);
		\draw [style={dw_and_one_hot_edge}] (15) to (1);
		\draw [style={dw_and_one_hot_edge}] (15) to (3);
		\draw [style={dw_and_one_hot_edge}] (16) to (2);
		\draw [style={dw_and_one_hot_edge}] (15) to (2);
		\draw [style={dw_and_one_hot_edge}] (16) to (3);
		\draw [style={dw_and_one_hot_edge}] (16) to (1);
		\draw [style={dw_and_one_hot_edge}] (16) to (12);
		\draw [style={dw_and_one_hot_edge}] (16) to (13);
		\draw [style={dw_and_one_hot_edge}] (16) to (11);
		\draw [style={one_hot_only_edge}] (18) to (2);
		\draw [style={one_hot_only_edge}] (18) to (3);
		\draw [style={one_hot_only_edge}] (18) to (1);
		\draw [style={one_hot_only_edge}] (18) to (12);
		\draw [style={one_hot_only_edge}] (18) to (11);
		\draw [style={one_hot_only_edge}] (13) to (18);
		\draw [style={one_hot_only_edge}] (10) to (15);
		\draw [style={one_hot_only_edge}] (10) to (17);
		\draw [style={one_hot_only_edge}] (10) to (16);
		\draw [style={one_hot_only_edge}] (14) to (17);
		\draw [style={one_hot_only_edge}] (14) to (15);
		\draw [style={one_hot_only_edge}] (14) to (16);
		\draw [style={one_hot_only_edge}] (10) to (18);
		\draw [style={one_hot_only_edge}] (14) to (18);
	\end{pgfonlayer}
\end{tikzpicture}
\par
\caption{Unstructured interactions between three $\mathbb{Z}_4$ variables. Black: qubits and interactions needed for both domain wall and one hot encoding. Magenta: additional qubits and interactions needed for one hot encoding. Edges within variables in the domain wall encoding have been made thicker as a guide to the eye.\label{fig:unstructured_one_hot_dw}}
\end{centering}
\end{figure}
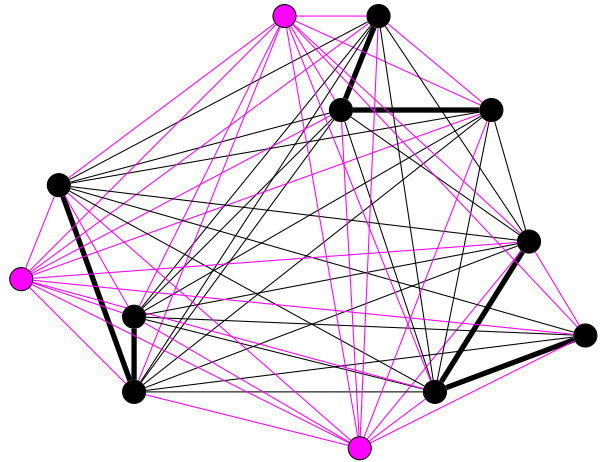

In the unstructured case, the interaction graph of the domain wall encoding will be a subgraph of the interaction graph of one hot, as depicted in Fig.~\ref{fig:unstructured_one_hot_dw}, therefore the domain wall encoding will always be easier to implement since all of the interactions needed for the domain wall encoding are also needed in one hot. In the example given in Fig.~\ref{fig:unstructured_one_hot_dw}, with unstructured interactions between three $\mathbb{Z}_4$ variables, the domain wall encoding requires nine qubits and $36$ interactions, while the one hot encoding requires $12$ qubits and $76$ interactions \footnote{By using a judicious choice of the zero of energy, this  could be reduced to $73$ interactions. This  `removed' edge can be placed between additional qubits in one hot relative to domain wall without loss of generality so the domain wall encoding graph remains a subgraph of the one hot graph}. Given that an advantage can be shown analytically (by showing that the domain wall encoding is a subgraph of the one hot), it is not necessary to numerically analyse unstructured problems to show an advantage.

\subsection{Graph Coloring\label{sub:coloring}}
Let us now consider the more structured problem of maximum graph colouring (referred to as Max-$\kappa$-ColorableSubgraph \cite{Hadfield17a}, also sometimes referred to as Max-$\kappa$-Cut \cite{Khot07a,Frieze97a}), where given a graph and $n$ possible node colours, the goal is to colour the graph to maximize the number of edges which connect vertexes of different colours.  Maximum graph colouring is a generalization of the more studied problem of graph colouring, since in graphs which are colourable with $n$ colours, the maximal colouring is a `proper' colouring of the graph, where no vertexes of the same colour share an edge. The question of whether or not a graph can be coloured is known to be NP-hard if the number of colours requires is greater than two \cite{Duffy2008a}, and even remains hard under quite restrictive conditions \cite{Dailey1980a}. Solutions to graph colouring problems have wide applicability, including in aircraft scheduling, organizing file transfer between processors, and radio frequency assignments \cite{Marx2004a}. Quantum annealing has been applied to graph colouring problems in \cite{Titiloye11a,Titiloye11b}.

The structure of the interactions for colouring problems is therefore to penalize vertexes of the same colour (variables which take the same value) while having no effect otherwise. Since this interaction maps directly to anti-ferromagnetic interactions between qubits corresponding to the same value in one hot, each edge in the colouring graph requires $n$ two qubit interaction. For the domain wall encoding, the interactions to enforce different colours are more complicated, but there is also one fewer qubit per variable. When $n\ge 3$, the number of interactions required per graph edge is $3\,(n-1)-2=3\,n-5$. This is not the end of the story when it comes to number of interactions, however, since the number of interactions per \emph{variable} is more for one hot, requiring $\frac{1}{2}n\,(n-1)$ edges compared to the $n-2$ interactions required by the domain wall encoding. 

\begin{figure}
\begin{centering}
\begin{tikzpicture}
	\begin{pgfonlayer}{nodelayer}
		\node [style={one_hot_node}] (0) at (2, 3) {};
		\node [style={one_hot_node}] (1) at (0.25, 0.75) {};
		\node [style={one_hot_node}] (2) at (4, 0.25) {};
		\node [style={one_hot_node}] (3) at (5.5, 3) {};
		\node [style={one_hot_node}] (4) at (2.5, 2.75) {};
		\node [style={one_hot_node}] (5) at (1.25, 0.25) {};
		\node [style={one_hot_node}] (6) at (4.75, -0.25) {};
		\node [style={one_hot_node}] (7) at (6, 2.75) {};
		\node [style={one_hot_node}] (8) at (1.75, 2.25) {};
		\node [style={one_hot_node}] (9) at (0.25, -0.25) {};
		\node [style={one_hot_node}] (10) at (4, -0.75) {};
		\node [style={one_hot_node}] (11) at (5.5, 2.25) {};
		\node [style={one_hot_node}] (12) at (1.25, 2.5) {};
		\node [style={one_hot_node}] (13) at (-0.5, 0.25) {};
		\node [style={one_hot_node}] (14) at (3.25, -0.25) {};
		\node [style={one_hot_node}] (15) at (4.75, 2.5) {};
		\node [style={coloring_node}] (16) at (5.75, 1.25) {};
		\node [style={coloring_node}] (17) at (5.25, 0.25) {};
		\node [style={coloring_node}] (18) at (6.5, -0.25) {};
		\node [style={coloring_node}] (19) at (7, 1.25) {};
	\end{pgfonlayer}
	\begin{pgfonlayer}{edgelayer}
		\draw [style={anti_ferro_coup}] (0) to (3);
		\draw [style={anti_ferro_coup}] (1) to (0);
		\draw [style={anti_ferro_coup}] (0) to (2);
		\draw [style={anti_ferro_coup}] (1) to (2);
		\draw [style={anti_ferro_coup}] (2) to (3);
		\draw [style={anti_ferro_coup}] (4) to (7);
		\draw [style={anti_ferro_coup}] (5) to (4);
		\draw [style={anti_ferro_coup}] (4) to (6);
		\draw [style={anti_ferro_coup}] (5) to (6);
		\draw [style={anti_ferro_coup}] (6) to (7);
		\draw [style={anti_ferro_coup}] (8) to (11);
		\draw [style={anti_ferro_coup}] (9) to (8);
		\draw [style={anti_ferro_coup}] (8) to (10);
		\draw [style={anti_ferro_coup}] (9) to (10);
		\draw [style={anti_ferro_coup}] (10) to (11);
		\draw [style={anti_ferro_coup}] (12) to (15);
		\draw [style={anti_ferro_coup}] (13) to (12);
		\draw [style={anti_ferro_coup}] (12) to (14);
		\draw [style={anti_ferro_coup}] (13) to (14);
		\draw [style={anti_ferro_coup}] (14) to (15);
		\draw [style={one_hot_edge}] (12) to (0);
		\draw [style={one_hot_edge}] (0) to (4);
		\draw [style={one_hot_edge}] (4) to (8);
		\draw [style={one_hot_edge}] (8) to (12);
		\draw [style={one_hot_edge}] (12) to (4);
		\draw [style={one_hot_edge}] (8) to (0);
		\draw [style={one_hot_edge}] (1) to (9);
		\draw [style={one_hot_edge}] (13) to (5);
		\draw [style={one_hot_edge}] (1) to (5);
		\draw [style={one_hot_edge}] (5) to (9);
		\draw [style={one_hot_edge}] (9) to (13);
		\draw [style={one_hot_edge}] (13) to (1);
		\draw [style={one_hot_edge}] (14) to (2);
		\draw [style={one_hot_edge}] (2) to (10);
		\draw [style={one_hot_edge}] (10) to (14);
		\draw [style={one_hot_edge}] (14) to (6);
		\draw [style={one_hot_edge}] (10) to (6);
		\draw [style={one_hot_edge}] (2) to (6);
		\draw [style={one_hot_edge}] (15) to (3);
		\draw [style={one_hot_edge}] (3) to (7);
		\draw [style={one_hot_edge}] (7) to (11);
		\draw [style={one_hot_edge}] (11) to (15);
		\draw [style={one_hot_edge}] (15) to (7);
		\draw [style={one_hot_edge}] (3) to (11);
		\draw [style={coloring_edge}] (16) to (19);
		\draw [style={coloring_edge}] (17) to (16);
		\draw [style={coloring_edge}] (16) to (18);
		\draw [style={coloring_edge}] (17) to (18);
		\draw [style={coloring_edge}] (18) to (19);
	\end{pgfonlayer}
\end{tikzpicture}
\par
\end{centering}
\caption{\label{fig:four_color_one_hot} One hot encoding of maximum four colouring of a small graph fragment, red edges indicate edges encoding interactions between variables, while green indicate the internal edges within each variable. The lower right of the figure depicts the graph fragment.}
\end{figure}
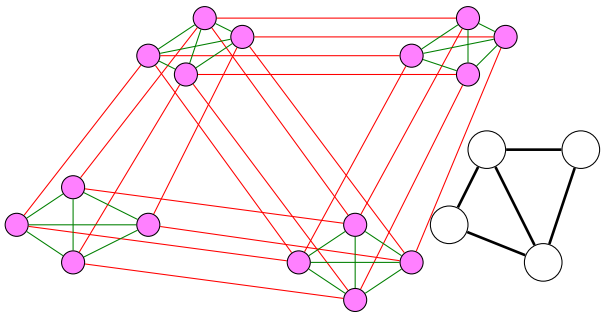

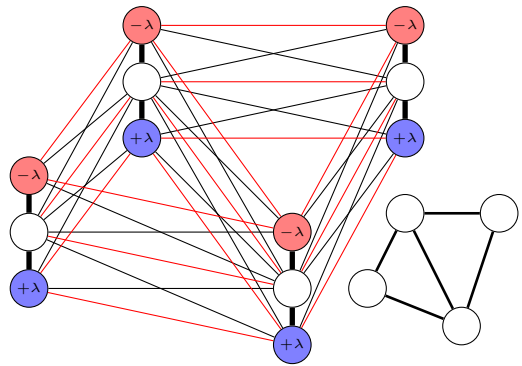
\begin{figure}
\begin{centering}
\begin{tikzpicture}
	\begin{pgfonlayer}{nodelayer}
		\node [style={plus_field_qubit}] (0) at (-1.75, 3.25) {};
		\node [style={plus_field_qubit}] (1) at (-3.25, 1.25) {};
		\node [style={plus_field_qubit}] (2) at (0.25, 0.5) {};
		\node [style={plus_field_qubit}] (3) at (1.75, 3.25) {};
		\node [style={no_field_qubit}] (4) at (-3.25, 0.5) {};
		\node [style={no_field_qubit}] (5) at (-1.75, 2.5) {};
		\node [style={no_field_qubit}] (6) at (1.75, 2.5) {};
		\node [style={no_field_qubit}] (7) at (0.25, -0.25) {};
		\node [style={minus_field_qubit}] (8) at (-3.25, -0.25) {};
		\node [style={minus_field_qubit}] (9) at (-1.75, 1.75) {};
		\node [style={minus_field_qubit}] (10) at (0.25, -1) {};
		\node [style={minus_field_qubit}] (11) at (1.75, 1.75) {};
		\node [style={coloring_node}] (12) at (1.75, 0.75) {};
		\node [style={coloring_node}] (13) at (1.25, -0.25) {};
		\node [style={coloring_node}] (14) at (2.5, -0.75) {};
		\node [style={coloring_node}] (15) at (3, 0.75) {};
	\end{pgfonlayer}
	\begin{pgfonlayer}{edgelayer}
		\draw [style={anti_ferro_coup}] (0) to (3);
		\draw [style={anti_ferro_coup}] (1) to (0);
		\draw [style={ferro_coupling}] (1) to (5);
		\draw [style={ferro_coupling}] (4) to (0);
		\draw [style={ferro_coupling}] (8) to (5);
		\draw [style={ferro_coupling}] (4) to (9);
		\draw [style={ferro_coupling}] (0) to (6);
		\draw [style={ferro_coupling}] (5) to (3);
		\draw [style={ferro_coupling}] (11) to (5);
		\draw [style={ferro_coupling}] (9) to (6);
		\draw [style={ferro_coupling}] (7) to (3);
		\draw [style={ferro_coupling}] (2) to (6);
		\draw [style={ferro_coupling}] (10) to (6);
		\draw [style={ferro_coupling}] (7) to (11);
		\draw [style={ferro_coupling}] (0) to (7);
		\draw [style={ferro_coupling}] (2) to (5);
		\draw [style={ferro_coupling}] (10) to (5);
		\draw [style={ferro_coupling}] (7) to (9);
		\draw [style={ferro_coupling}] (2) to (4);
		\draw [style={ferro_coupling}] (7) to (1);
		\draw [style={ferro_coupling}] (8) to (7);
		\draw [style={ferro_coupling}] (10) to (4);
		\draw [style={anti_ferro_coup}] (4) to (5);
		\draw [style={anti_ferro_coup}] (5) to (6);
		\draw [style={anti_ferro_coup}] (6) to (7);
		\draw [style={anti_ferro_coup}] (7) to (4);
		\draw [style={anti_ferro_coup}] (7) to (5);
		\draw [style={anti_ferro_coup}] (10) to (9);
		\draw [style={anti_ferro_coup}] (10) to (11);
		\draw [style={anti_ferro_coup}] (10) to (8);
		\draw [style={anti_ferro_coup}] (8) to (9);
		\draw [style={anti_ferro_coup}] (9) to (11);
		\draw [style={anti_ferro_coup}] (0) to (2);
		\draw [style={anti_ferro_coup}] (1) to (2);
		\draw [style={anti_ferro_coup}] (2) to (3);
		\draw [style={domain_wall_edge}] (0) to (5);
		\draw [style={domain_wall_edge}] (5) to (9);
		\draw [style={domain_wall_edge}] (1) to (4);
		\draw [style={domain_wall_edge}] (4) to (8);
		\draw [style={domain_wall_edge}] (2) to (7);
		\draw [style={domain_wall_edge}] (7) to (10);
		\draw [style={domain_wall_edge}] (3) to (6);
		\draw [style={domain_wall_edge}] (6) to (11);
		\draw [style={coloring_edge}] (12) to (15);
		\draw [style={coloring_edge}] (13) to (12);
		\draw [style={coloring_edge}] (12) to (14);
		\draw [style={coloring_edge}] (13) to (14);
		\draw [style={coloring_edge}] (14) to (15);
	\end{pgfonlayer}
\end{tikzpicture}
\par
\end{centering}
\caption{\label{fig:four_color_domain_wall}Domain wall encoding of maximum four colouring of a small graph fragment, red edges indicate anti-ferromagnetic two qubit interactions between variables, while black edges indicate the same but ferromagnetic. Thick black the internal edges within each variable. Red and blue qubits indicate where single body terms are applied, following Fig.~\ref{fig:domain_wall_virtual}(lower). The lower right of the figure depicts the graph fragment.}
\end{figure}

As an example, I show how to encode maximum four colouring on this four qubit graph fragment fig.~\ref{fig:four_color_one_hot} depicts the one hot encoding for maximum four colouring on a four vertex graph fragment, while Fig.~\ref{fig:four_color_domain_wall} depicts the domain wall encoding.

For $n$ colours, it is possible calculate the ratio $r$ of edges to variables above which one hot will involve fewer interactions, and below which the domain wall encoding will require fewer.  This calculation is performed by first finding the number of interactions per vertex required in the one hot and domain wall cases, and then setting them equal and solving. For the one hot encoding, each vertex will require $\frac{1}{2}n\,(n-1)$ internal interactions, and each edge will require another $n$. For a given $r$ ratio of edges to vertexes, this means that there will be $n\,(n-1)+r\,n$ interactions per vertex. On the other hand, for the domain wall encoding, each vertex requires only $n-1$ internal interactions, but each edge requires $3\,n+1$ interactions, leading to a total of $n-1+3\,r \,n+r$. Setting the two expressions equal and solving for $r$ leads to:
\begin{equation}
r_c(n)=\frac{\frac{1}{2}n^2-\frac{3}{2}\,n+2}{2\,n-5}
\end{equation}
 assuming again $n\ge 3$. In the limit of large $n$ this expression goes as $\frac{n}{4}$, considering that each vertex should be adjacent to at least $n$ other vertexes for the colouring problem to be non-trivial, for a large number of colours the domain wall encoding will contain more edges for realistic problems. For a smaller number of colours, this ratio can be larger for instance $r_c(3)=2$, and $r_c(4)=\frac{4}{3}$. It is important to recall that in all cases, the domain wall encoding requires fewer qubits. There are also important differences in the structure of the interaction, which will be highlighted in the next section.

Ignoring the differences in interaction graph structure, in cases where $r>r_c$ there is a tradeoff in terms of interaction number versus qubit number, with the domain wall encoding requiring fewer qubits but more interactions. Although a gross oversimplification, let us consider for a moment the case where we ignore the structure and consider interaction counts. This over simplified picture suggest that for instance in an optical setting \cite{Inagaki16a,McMahon16a} than the domain wall encoding would be preferred. However, if interactions are more difficult to implement, then the one hot encoding may be best.  As I demonstrate later, the opposite is actually true , the structure of the domain wall encoding makes it easier to implement, and therefore also preferable at large sizes.

In real situations, it is not just the number of edges which is important, but also the structure of the edges.  We first note that the domain wall encoding forms a `layered' structure, where the qubits can be divided into $n-1$ layers corresponding to position in the chain used for the domain wall encoding. The minimum possible edge distance $d_e$ between a qubit on layer $i$ and one on layer $j$ is $|i-j|$, regardless of the graph being coloured, such a structure is not present in the one hot encoding. As discussed later, numerical analysis reveals that in fact the relative ease of embedding the structures makes domain wall similarly or more efficient to embed for all problem sizes I analyse.

To examine the effect of the different structures, I consider numerically embedding maximum $n$ colour problems on Erd\"os-R\'enyi random graphs \cite{erdos59a,erdos60a}(each pair of vertices independently has an edge with an independent fixed probability)  with edge probabilities of $0.75$ and $2\, n$ vertices. As a comparison, I also examine maximum three colour problems on Erd\"os-R\'enyi random graphs with edge probabilities of $0.5$.  I examine embeddings on both the D-Wave Chimera graph and the recently proposed Pegasus \cite{Boothby19a} graph. 

For this analysis, I find the smallest Pegasus or square Chimera graph for which a given problem can be embedded, using the available software \cite{minorminer}. I refer to either the linear size of the Chimera (number of unit cells along one side) or the size of the Pegasus graph (encoded in a single number) as $L$. The exact methods which are used for the numerics are described in Sec.~\ref{sec:methods}.

\begin{figure}
\begin{centering}
\includegraphics[width=7 cm]{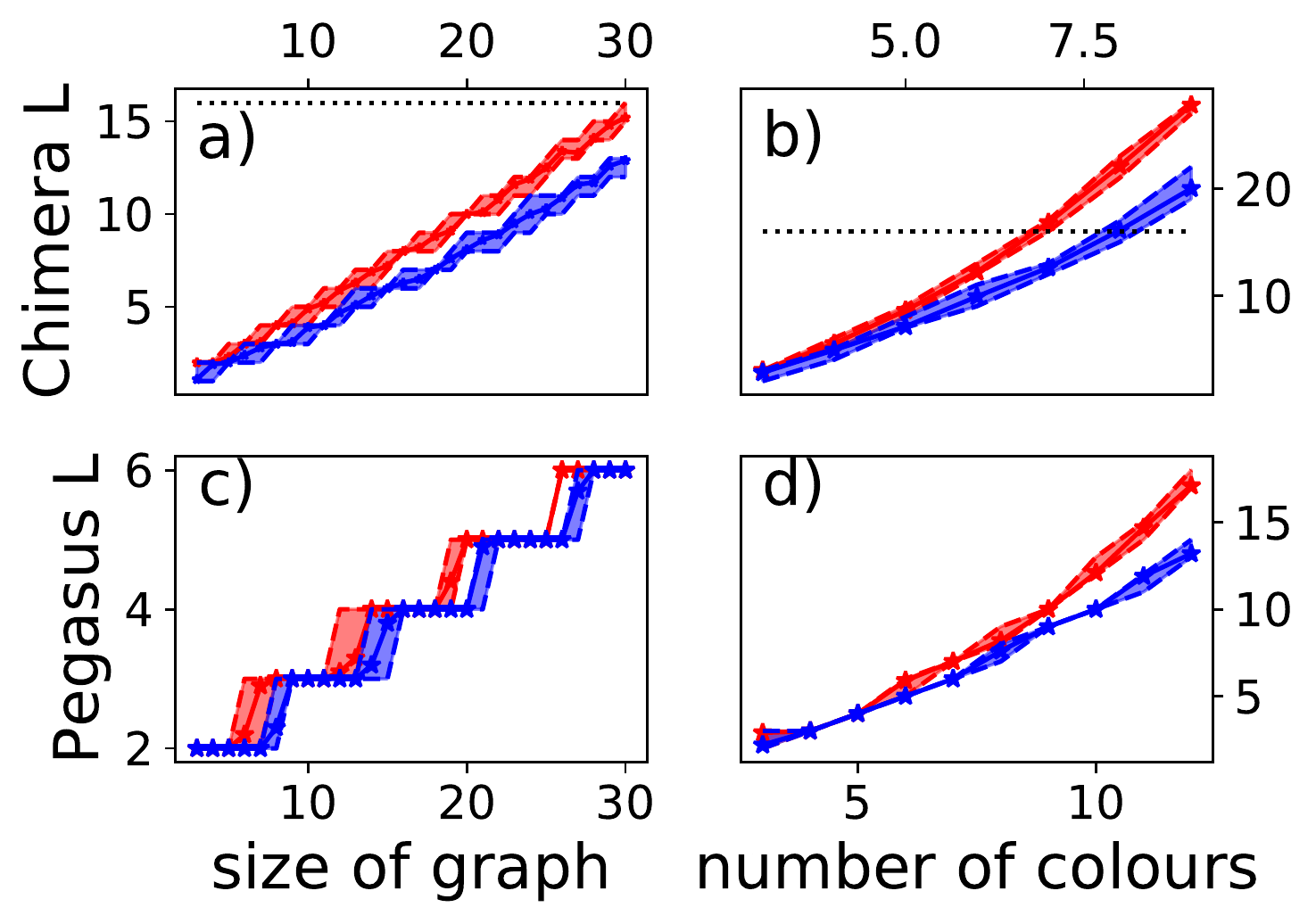}
\par
\end{centering}
\caption{\label{fig:min_embedding_size_color}Minimum linear size of graph requires to embed one hot (red) and domain wall (blue) encoded colouring problems. Dashed lines and shaded regions indicate minimum and maximum seen for $10$ instances and the solid line with stars indicates the average, dotted lines on Chimera plots indicate size of the current generation of D-Wave device. (a) Three color embedded into Chimera. (b) $n$ color embedding into chimera. (c) Three colour embedding into Pegasus. (d) $n$ colour embedding into Pegasus.}
\end{figure}

As we can see in Fig.~\ref{fig:min_embedding_size_color}, the minimum size of chimera or Pegasus graph where a problem can successfully be embedded is always smaller or equal on average for the domain wall encoding versus one hot encoding. Moreover, except for the three colour embedding in Pegasus at large sizes the difference becomes more dramatic, and the worst embedding of a domain wall encoding is still superior to the best for one hot. Finally, we observe that while, for the three colour problems the difference between the domain wall and one hot encodings is minimal and grows only slowly with size, it grows much more dramatically for the $n$ colour problem. As I demonstrate later, in subsection \ref{sub:analysis}, this is due to the domain wall encoding being more efficiently embeddable, likely because of the previously mentioned layered structure of the domain wall encoding in the $n$ colour case.

\subsection{Scheduling}

Let us now consider the problem of minimizing (or eliminating) scheduling conflicts, different versions of this problem have been considered for quantum computing \cite{Venturelli15a,Stollenwerk19a,crispin13a,Tran16a}, including most recently the problem of flight deconflicting \cite{Stollenwerk19a}. The basic structure I consider is that there are $N_t$ possible times and  $m$ events each of duration $T_k$ each of which which must start at time $t_k \in (t_{k,\mathrm{min}},t_{k,\mathrm{max}})$ where $t_{k,\mathrm{min}}\in (0,N-1)$ and $t_{k,\mathrm{max}}\ge t_0 \in (0,N-1-T_k)$. Moreover, conflicts can occur if certain pairs of events, occuring at $t_k$ and $t_l$ overlap, which means that either $0\le t_l-t_k<T_k$ or $0\le t_k-t_l<T_l$. In both the domain wall and one hot encoding, single variable penalties correspond to single body Ising terms in the encoding, therefore, the problem structure is not changed by adding such penalties, which could correspond for example to penalties for delaying a flight in \cite{Stollenwerk19a}.

\begin{figure}
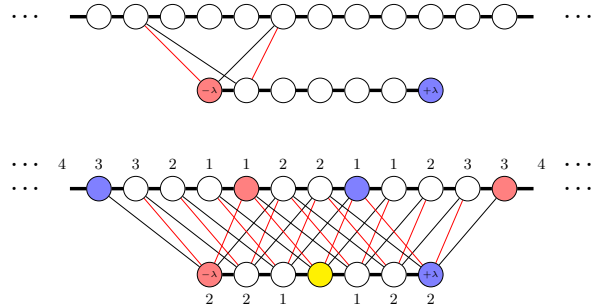

\input{tikzit/scheduling_interaction.label.tex} 
\begin{centering}
\resizebox{8 cm}{!}{\begin{tikzpicture}
	\begin{pgfonlayer}{nodelayer}
		\node [style={no_field_qubit}] (4) at (0, 4.5) {};
		\node [style={no_field_qubit}] (5) at (0.75, 4.5) {};
		\node [style={no_field_qubit}] (6) at (1.5, 4.5) {};
		\node [style={no_field_qubit}] (7) at (2.25, 4.5) {};
		\node [style={no_field_qubit}] (8) at (3, 4.5) {};
		\node [style={no_field_qubit}] (9) at (3.75, 4.5) {};
		\node [style={no_field_qubit}] (10) at (4.5, 4.5) {};
		\node [style={no_field_qubit}] (11) at (5.25, 4.5) {};
		\node [style={no_field_qubit}] (12) at (6, 4.5) {};
		\node [style={edge_end}] (13) at (6.75, 4.5) {};
		\node [style={plus_field_qubit}] (14) at (0, 3) {};
		\node [style={no_field_qubit}] (15) at (0.75, 3) {};
		\node [style={no_field_qubit}] (16) at (1.5, 3) {};
		\node [style={no_field_qubit}] (17) at (2.25, 3) {};
		\node [style={no_field_qubit}] (18) at (3, 3) {};
		\node [style={no_field_qubit}] (19) at (3.75, 3) {};
		\node [style={no_field_qubit}] (24) at (-0.75, 4.5) {};
		\node [style={no_field_qubit}] (25) at (-1.5, 4.5) {};
		\node [style={no_field_qubit}] (26) at (-2.25, 4.5) {};
		\node [style={minus_field_qubit}] (29) at (4.5, 3) {};
		\node [style={dot_dot_dot}] (31) at (-3.75, 4.5) {};
		\node [style={dot_dot_dot}] (32) at (7.5, 4.5) {};
		\node [style={edge_end}] (33) at (-3, 4.5) {};
		\node [style={no_field_qubit}] (34) at (0, 1) {};
		\node [style={plus_filed_nolabel}] (35) at (0.75, 1) {};
		\node [style={no_field_qubit}] (36) at (1.5, 1) {};
		\node [style={no_field_qubit}] (37) at (2.25, 1) {};
		\node [style={minus_field_nolabel}] (38) at (3, 1) {};
		\node [style={no_field_qubit}] (39) at (3.75, 1) {};
		\node [style={no_field_qubit}] (40) at (4.5, 1) {};
		\node [style={no_field_qubit}] (41) at (5.25, 1) {};
		\node [style={plus_filed_nolabel}] (42) at (6, 1) {};
		\node [style={edge_end}] (43) at (6.75, 1) {};
		\node [style={plus_field_qubit}] (44) at (0, -0.75) {};
		\node [style={no_field_qubit}] (45) at (0.75, -0.75) {};
		\node [style={no_field_qubit}] (46) at (1.5, -0.75) {};
		\node [style={distance_node}] (47) at (2.25, -0.75) {};
		\node [style={no_field_qubit}] (48) at (3, -0.75) {};
		\node [style={no_field_qubit}] (49) at (3.75, -0.75) {};
		\node [style={minus_field_qubit}] (50) at (4.5, -0.75) {};
		\node [style={no_field_qubit}] (51) at (-0.75, 1) {};
		\node [style={no_field_qubit}] (52) at (-1.5, 1) {};
		\node [style={minus_field_nolabel}] (53) at (-2.25, 1) {};
		\node [style={dot_dot_dot}] (57) at (-3.75, 1) {};
		\node [style={dot_dot_dot}] (58) at (7.5, 1) {};
		\node [style={edge_end}] (59) at (-3, 1) {};
		\node [style={label_2}] (60) at (1.5, 1.5) {};
		\node [style={label_2}] (61) at (2.25, 1.5) {};
		\node [style={label_1}] (62) at (3, 1.5) {};
		\node [style={label_1}] (63) at (0.75, 1.5) {};
		\node [style={label_1}] (64) at (1.5, -1.25) {};
		\node [style={label_1}] (65) at (3, -1.25) {};
		\node [style={label_1}] (66) at (3.75, 1.5) {};
		\node [style={label_2}] (67) at (-0.75, 1.5) {};
		\node [style={label_3}] (68) at (-1.5, 1.5) {};
		\node [style={label_3}] (69) at (-2.25, 1.5) {};
		\node [style={label_1}] (70) at (0, 1.5) {};
		\node [style={label_2}] (71) at (4.5, 1.5) {};
		\node [style={label_3}] (72) at (5.25, 1.5) {};
		\node [style={label_3}] (73) at (6, 1.5) {};
		\node [style={label_2}] (74) at (0.75, -1.25) {};
		\node [style={label_2}] (75) at (3.75, -1.25) {};
		\node [style={label_2}] (76) at (4.5, -1.25) {};
		\node [style={label_2}] (77) at (0, -1.25) {};
		\node [style={label_4}] (78) at (-3, 1.5) {};
		\node [style={label_4}] (80) at (6.75, 1.5) {};
		\node [style={dot_dot_dot}] (81) at (7.5, 1.5) {};
		\node [style={dot_dot_dot}] (82) at (-3.75, 1.5) {};
	\end{pgfonlayer}
	\begin{pgfonlayer}{edgelayer}
		\draw [style={domain_wall_edge}] (4) to (5);
		\draw [style={domain_wall_edge}] (5) to (6);
		\draw [style={domain_wall_edge}] (6) to (7);
		\draw [style={domain_wall_edge}] (7) to (8);
		\draw [style={domain_wall_edge}] (8) to (9);
		\draw [style={domain_wall_edge}] (9) to (10);
		\draw [style={domain_wall_edge}] (10) to (11);
		\draw [style={domain_wall_edge}] (11) to (12);
		\draw [style={domain_wall_edge}] (12) to (13);
		\draw [style={domain_wall_edge}] (14) to (15);
		\draw [style={domain_wall_edge}] (15) to (16);
		\draw [style={domain_wall_edge}] (16) to (17);
		\draw [style={domain_wall_edge}] (17) to (18);
		\draw [style={domain_wall_edge}] (18) to (19);
		\draw [style={domain_wall_edge}] (26) to (25);
		\draw [style={domain_wall_edge}] (25) to (24);
		\draw [style={domain_wall_edge}] (24) to (4);
		\draw [style={anti_ferro_coup}] (14) to (25);
		\draw [style={anti_ferro_coup}] (15) to (6);
		\draw [style={ferro_coupling}] (14) to (6);
		\draw [style={ferro_coupling}] (15) to (25);
		\draw [style={domain_wall_edge}] (33) to (26);
		\draw [style={domain_wall_edge}] (34) to (35);
		\draw [style={domain_wall_edge}] (35) to (36);
		\draw [style={domain_wall_edge}] (36) to (37);
		\draw [style={domain_wall_edge}] (37) to (38);
		\draw [style={domain_wall_edge}] (38) to (39);
		\draw [style={domain_wall_edge}] (39) to (40);
		\draw [style={domain_wall_edge}] (40) to (41);
		\draw [style={domain_wall_edge}] (41) to (42);
		\draw [style={domain_wall_edge}] (42) to (43);
		\draw [style={domain_wall_edge}] (44) to (45);
		\draw [style={domain_wall_edge}] (45) to (46);
		\draw [style={domain_wall_edge}] (46) to (47);
		\draw [style={domain_wall_edge}] (47) to (48);
		\draw [style={domain_wall_edge}] (48) to (49);
		\draw [style={domain_wall_edge}] (49) to (50);
		\draw [style={domain_wall_edge}] (53) to (52);
		\draw [style={domain_wall_edge}] (52) to (51);
		\draw [style={domain_wall_edge}] (51) to (34);
		\draw [style={anti_ferro_coup}] (44) to (52);
		\draw [style={anti_ferro_coup}] (45) to (36);
		\draw [style={ferro_coupling}] (44) to (36);
		\draw [style={ferro_coupling}] (45) to (52);
		\draw [style={domain_wall_edge}] (59) to (53);
		\draw [style={ferro_coupling}] (44) to (53);
		\draw [style={anti_ferro_coup}] (44) to (35);
		\draw [style={anti_ferro_coup}] (45) to (51);
		\draw [style={ferro_coupling}] (46) to (51);
		\draw [style={anti_ferro_coup}] (46) to (37);
		\draw [style={ferro_coupling}] (45) to (37);
		\draw [style={anti_ferro_coup}] (46) to (34);
		\draw [style={anti_ferro_coup}] (47) to (38);
		\draw [style={ferro_coupling}] (46) to (38);
		\draw [style={ferro_coupling}] (47) to (34);
		\draw [style={anti_ferro_coup}] (47) to (35);
		\draw [style={anti_ferro_coup}] (48) to (39);
		\draw [style={ferro_coupling}] (47) to (39);
		\draw [style={ferro_coupling}] (48) to (35);
		\draw [style={anti_ferro_coup}] (48) to (36);
		\draw [style={domain_wall_edge}] (19) to (29);
		\draw [style={anti_ferro_coup}] (49) to (40);
		\draw [style={ferro_coupling}] (48) to (40);
		\draw [style={ferro_coupling}] (49) to (36);
		\draw [style={anti_ferro_coup}] (49) to (37);
		\draw [style={anti_ferro_coup}] (50) to (41);
		\draw [style={ferro_coupling}] (49) to (41);
		\draw [style={ferro_coupling}] (50) to (37);
		\draw [style={anti_ferro_coup}] (50) to (38);
		\draw [style={ferro_coupling}] (50) to (42);
	\end{pgfonlayer}
\end{tikzpicture}}
\par
\input{tikzit/blank.label.tex} 
\end{centering}
\caption{\label{scheduling_interaction} Encoding of scheduling conflicts using domain wall variables. Top: the encoding of a conflicts at a single time, where the event encoded in the the top chain has a duration of three time units and the event encoded in the bottom has a duration of two. Bottom: total encoding of the conflict with labels indicating the edge distance from the yellow vertex, the `...' indicates a continued linear increase in edge distance. Otherwise, colours have the same meaning as in Fig.~\ref{fig:domain_wall_virtual}(lower).}
\end{figure}

The structure for encoding time conflicts into the domain wall encoding can be found in Fig.~\ref{scheduling_interaction}(top) we see the encoding of all conflicts which could occur if the lower variable has the value corresponding to the first domain wall position on the bottom variable. In this figure the duration of the top domain wall encoded variable is $T_k=3$ time units, whereas the duration of the event encoded in the bottom vairable is $T_l=2$. To encode the total conflicts, we just add encodings for the conflicts at all allowed values of $t_l$, the result is Fig.~\ref{scheduling_interaction}(bottom). We also note the edge distance $d_e$ of all of the vertexes from the yellow vertex, in particular noting that continuing the figure beyond what is drawn, for the top variable the edge distance $d_e$ continues to grow linearly. In contrast, $d_e$ takes a maximum value of two for the one hot encoding regardless of the allowed time range of the two events (one hot encoding not shown).

For each possible value of a $t_l$ where there is a potential conflict with event $k$ and none of the domain wall variable values involved are the maximal or minimum possible values the conflicts can be encoded using four interactions between pairs of binary variables, if any of the variables do take extremal values, than the number of pairwise interactions will be less than four. It follows that if there are $q$ potential values where a conflict is possible, than the number of binary interactions which are needed is \emph{at most} $4\,q$, independent of the durations of each event, $T_k$ and $T_l$. For one hot encoding on the other hand, for every pair of times where there is a conflict, there must be an interaction between two binary variables, therefore the number of interactions grows with event duration.

\begin{figure}
\begin{centering}
\includegraphics[width=7 cm]{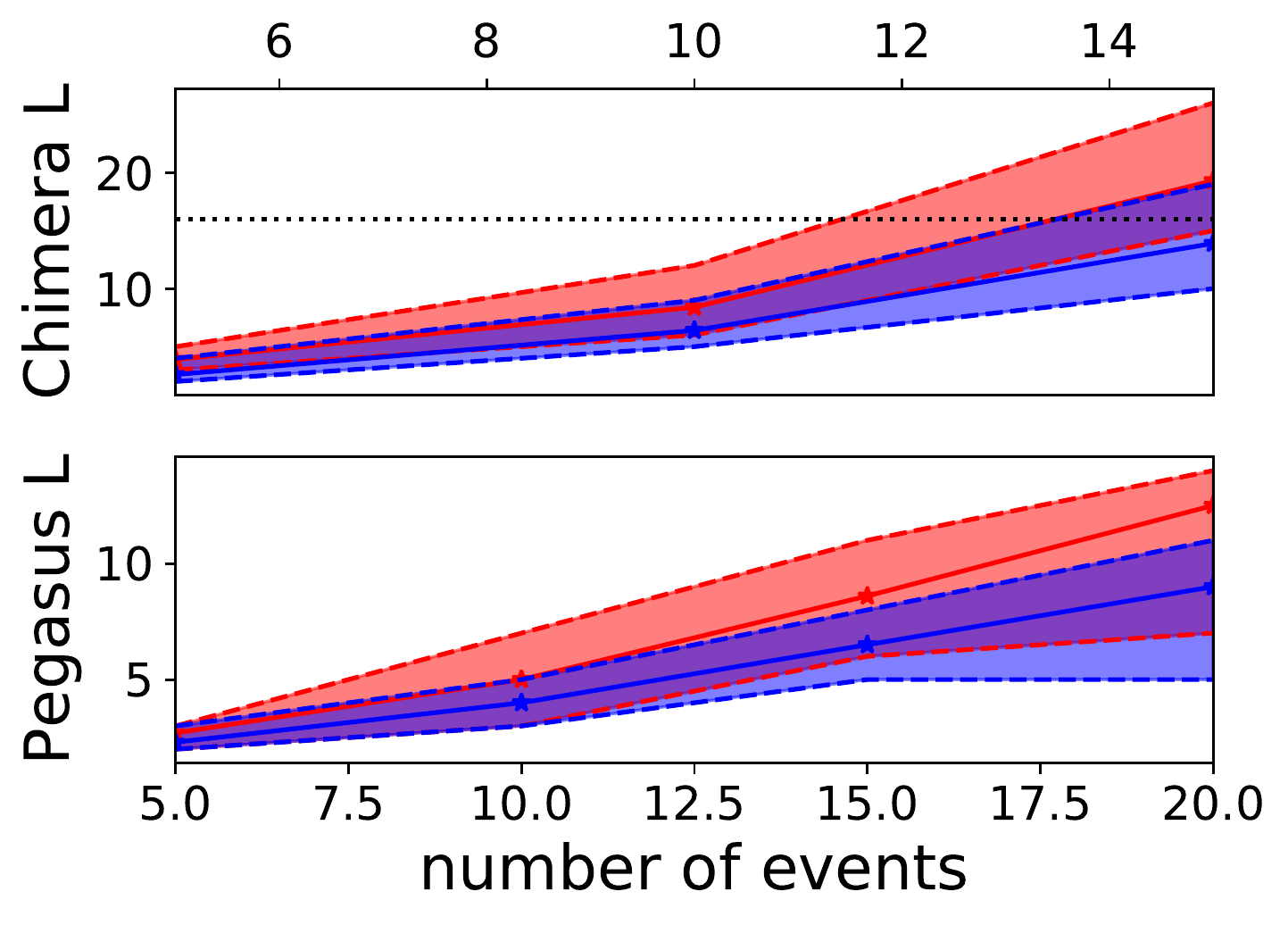}
\par
\end{centering}
\caption{\label{fig:min_embedding_size_sched}Minimum linear size of graph requires to embed one hot (red) and domain wall (blue) encoded schedule conflict minimization problems with different numbers events. Dashed lines and shaded regions indicate minimum and maximum seen for $10$ instances and the solid line with stars indicates the average, dotted lines on Chimera plots indicate size of the current generation of D-Wave device. top: Chimera, bottom: Pegasus}
\end{figure}

To get a sense of the effect of domain wall encoding, I again consider examples of embedding the same problem numerically when encoded in one hot versus domain wall encoding strategies. In this case, I consider random schedule conflict minimization problems using a construction discussed in detail in Sec.~\ref{sec:methods} where both the number of variables and the potential range of times are increased as the problems are scaled. The results of this embedding are depicted in Fig.~\ref{fig:min_embedding_size_sched}, while the range of graph sizes need to embed is much larger than for the case of colouring problems, the average size required for domain wall encodings is still significantly smaller. As I show in the next subsection, this difference is at least partially due to the domain wall encoding being easier to embed.

\subsection{Analysis\label{sub:analysis}}

So far I have demonstated the advantage of domain wall encodings in realistic problems but have not analyzed the source of the advantage, or compared relative advantages when embedding different problems into graphs. To do this, I define the \emph{embedding ratio}, which is the ratio of the number of vertexes used in the graph embedding to the number of vertexes in the original interaction graph. The embedding ratio captures the efficiency with which the problem structure can be embedded in a way which does not directly depend on the number of vertexes in the interaction graph, or details of the original problem.

\begin{figure}
\begin{centering}
\includegraphics[width=7 cm]{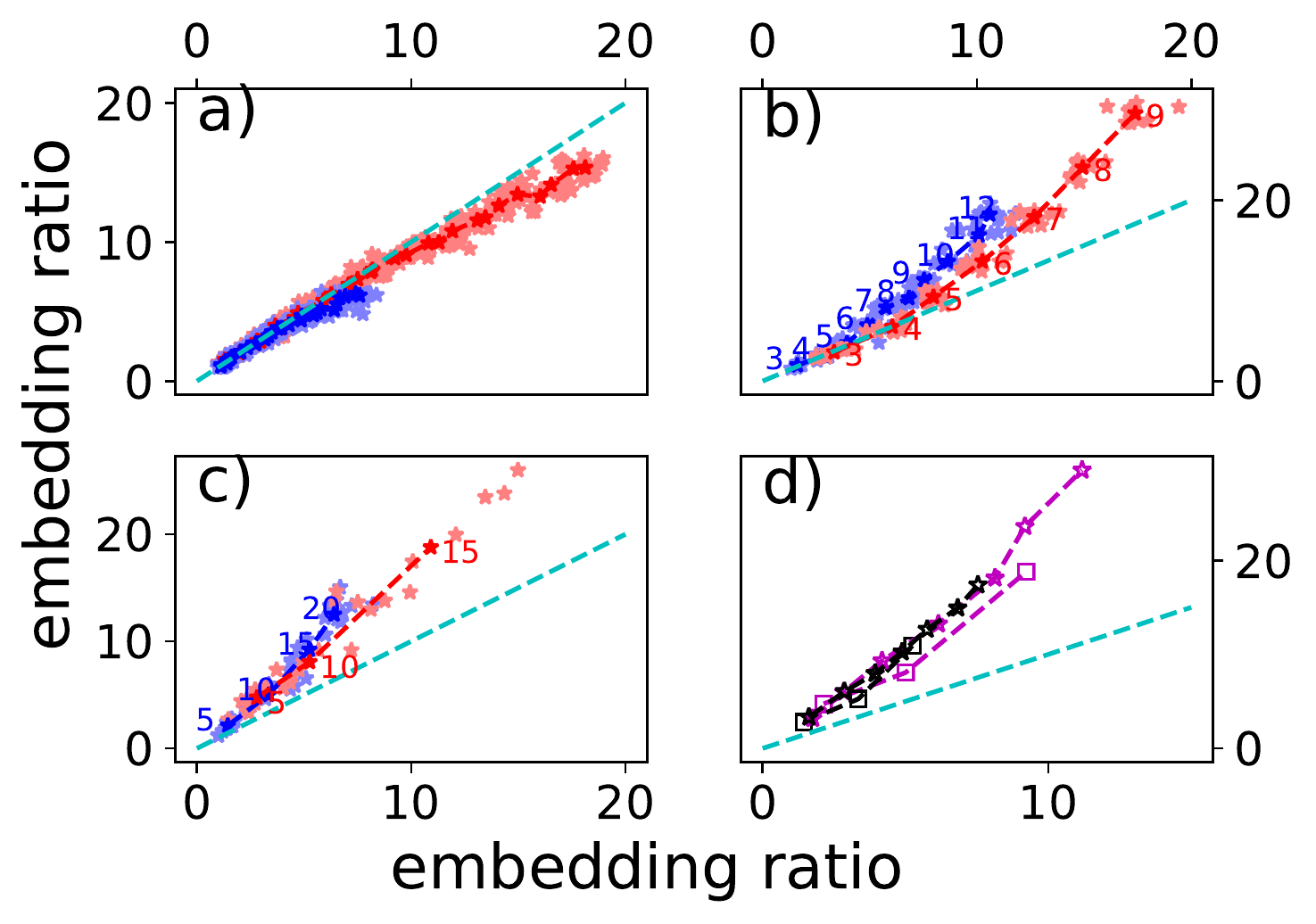}
\par
\end{centering}
\caption{\label{fig:multi_frame_embedding_ratio} Relative embedding ratios under different circumstances. Solid symbols (red and blue) indicate comparisons of domain wall encodings on the $X$ axis and one hot encodings on the $Y$ axis. Hollow symbols indicate embedding into the Pegasus graph on the $X$ axis and chimera on the $Y$. Red indicates embedding into a chimera graph, while blue indicates embedding into a Pegasus. For hollow symbols, black indicates one hot encoding, and magenta indicates domain wall. Fully coloured symbols indicate average values (over $10$ instances each), while lighter symbols indicate individual instances. All dashed lines are guides to the eye, with the cyan dashed line indicating circumstances where equal embedding ratios are obtained. (a) Three colour problems for sizes up to $30$ vertices. (b) $n$ colour problems, colour coded numbers indicate number of colours in the problem (c) Scheduling problems coloured numbers indicate number of events in the problem (d) Squares indicate scheduling problems, while stars indicate $n$ colour problems (three colour problems not shown).}
\end{figure}

Fig.~\ref{fig:multi_frame_embedding_ratio} compares the embedding ratios for the example problems discussed earlier in this section. From Fig.~\ref{fig:multi_frame_embedding_ratio}(a) it can be observed that the domain wall encoding of the maximum three colour problem is actually slightly less efficient to embed than the one hot encoding, therefore, the advantage seen in Fig.~\ref{fig:min_embedding_size_color}(a and c) is entirely because of the fact that the domain wall encoding requires fewer variables. However, Fig.~\ref{fig:multi_frame_embedding_ratio}(b and c) demonstrate that for the max $n$ color and scheduling problems the structure of the domain wall encoding can be embedded much more efficiently at large sizes. Finally, Fig.~\ref{fig:multi_frame_embedding_ratio}(d) demonstrates the advantage of embedding into a Pegasus rather than Chimera graph, with the Pegasus embedding being much more efficient.

\begin{figure}
\begin{centering}
\includegraphics[width=7 cm]{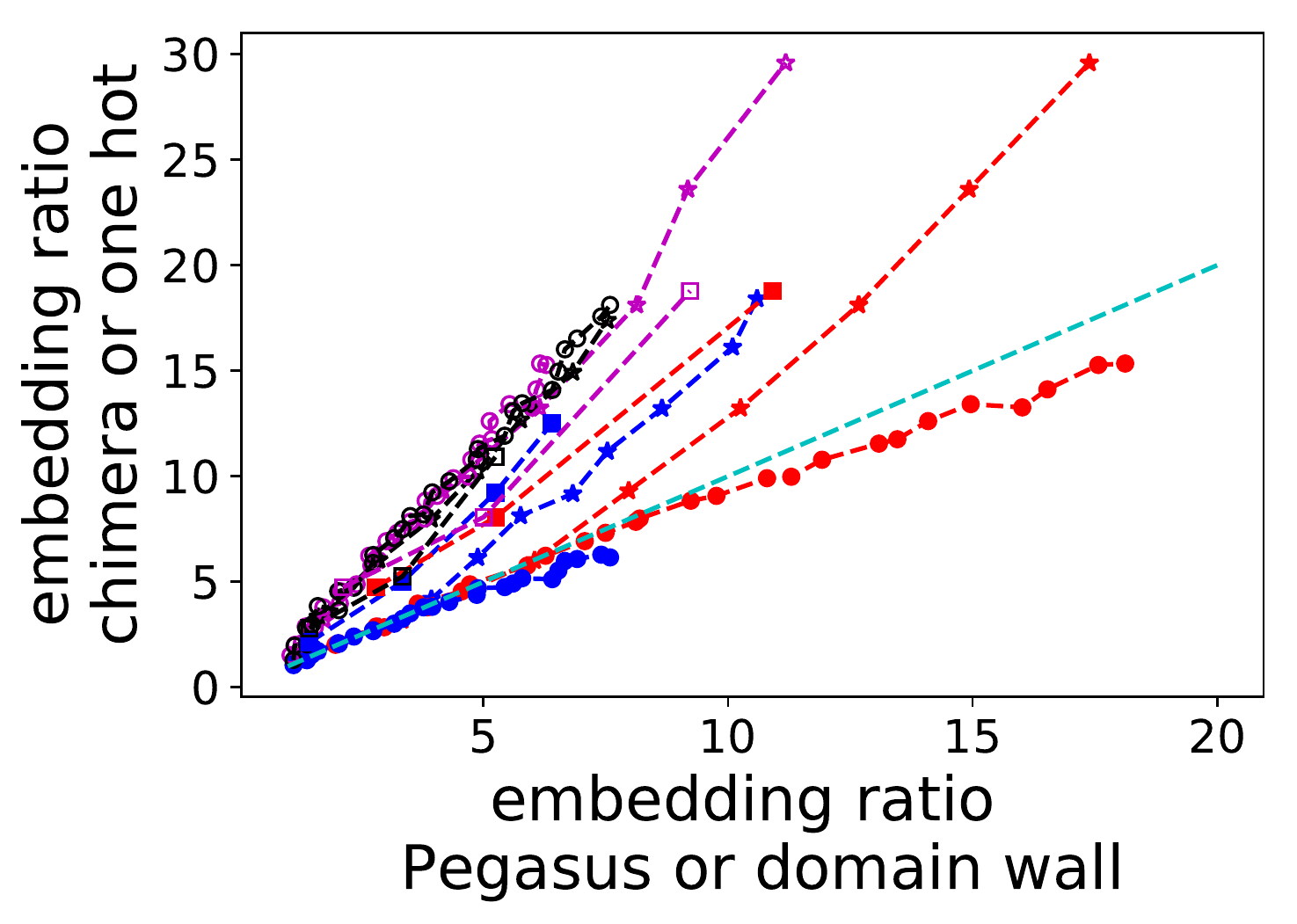}
\par
\end{centering}
\caption{\label{fig:single_plot_embedding_ratio} Relative embedding ratios under different circumstances. Solid symbols (red and blue) indicate domain wall encoding embedding ratio on the $X$ axis and one hot on the $Y$, whereas hollow symbols (black and magenta) indicate embedding into Pegasus on the $X$ versus embedding into Chimera on the $Y$. Red indicates embedding into a Chimera graph, while blue indicates embedding into a Pegasus. Black indicates one hot encoding, while Magenta indicates domain wall encoding. Circles indicate three colour problems, while stars indicate, $n$ colour problems, and squares indicate scheduling. All dashed lines are guides to the eye, with the cyan dashed line indicating the point where equal embedding ratios are obtained. All points are averaged over $10$ instances.}
\end{figure}

One additional advantage of comparing embedding ratios is that it provides a method to compare problems of different sizes and types, and even the relative advantages gained from changing different aspects, for instance encoding and hardware graph. Fig.~\ref{fig:single_plot_embedding_ratio} depicts relative embedding ratios under different circumstances plotted on the same axes. From this plot, we first observe that the structural advantage of domain wall encoding over one hot encoding is highly problem structure dependent, from actually a slight disadvantage in the case of three colour problems, to an advantage comparable with the advantage gained from embedding into a Pegasus graph rather than a chimera in the case of scheduling problems embedded into the Pegasus graph. The hardware graph structure (Pegasus versus chimera) on the other hand yields a consistent large advantage in terms of embedding overhead for all studied problem structure. 

\section{Numerical Methods \label{sec:methods}}

All embedding was performed using the minorminer software which is publicly available  \cite{minorminer}, with the default settings. Pegasus and Chimera graphs were created using the publicly available D-Wave networkx software \cite{D-Wave_nx}. All of the code for the numerical calculations was written in Python $3.5$ and is publicly available from \cite{domain_wall_code}. 

The minimum embeddable size for a given problem was calculated by first trying the size of the previous problem (or minimum size for the graph type) if embedding fails then the graph size is incremented until success and if successful the size is decremented until failure. For maximum colouring problems, the graphs to be coloured are Erd\"os R\'eyni random graphs \cite{erdos59a,erdos60a} (each pair of vertices independently has an edge with an independent fixed probability) with edge probability $0.5$ in the three colour case and $0.75$ in the $n$ colour case. 

For scheduling problems, the goal is to minimize the number of conflicts between events each of which is constrained to occur at integer times within a range of times between $0$ and $t_{max}$. The value of $t_{max}$ is chosen to be two times the number of events, and the probability that each pair of events will conflict (i.e.~that interactions need to be encoded between them) is chosen independently at random with the probability of a conflict being $0.75$. The earliest possible start time $t_{early}$ of an event is chosen uniformly at random between zero and $t_{max}-2$.  The latest possible start time $t_{late}$ is chosen uniformly at random between $t_{early}+1$ and $t_{max}$. The duration of each event is chosen uniformly at random between one and five time units.

The problems selected here have not been chosen to be provably hard, but rather to have representative structure of a problem type. It is, however, worth noting that even if the scheduling and colouring problem types which are use here are not asymptotically hard, there will be a plethora of weighted versions of the problems which will have the same interaction graphs (and therefore use the same embeddings), and it is likely that at least one of these versions will be hard. The importance of the existence of multiple problems with the same graph structure has been highlighted in \cite{Abbott18a}.

\section{Extension: Domain wall analogue of $k$-hot encodings \label{sec:k-hot}}

One additional advantage of a one-hot encoding is that it can be naturally extended to a $k$-hot encoding by modifying the strength of the one body terms such that in the lowest energy manifold $k$ variables are in the $1$ state, rather than only one. Unfortunately, it is not possible to play a similar trick for domain wall variables; the lowest energy state of the chain will always be the state with exactly one domain wall. The domain wall encoding can however be used to produce a $k$-hot analogy by linking together multiple chains and introducing strong interactions which do not allow any domain walls to be at the same site number. One way to accomplish this is employ the colouring problem encoding in section \ref{sub:coloring} on a clique (fully connected) graph, this would enforce that no two variables take the same value and thus the collective object behaves like a k-hot encoding.  

However, an analogue of the $k$-hot encoding requiring even less interaction between the chains is possible, this can come by realizing that the constraint that the domain wall variable $j+1$ has a greater value than variable $j$ can be implemented efficiently by only interacting nearby qubits on neighbouring chains.

This can be achieved by realising that iff the value of variable $j+1$ is less than or equal to that of variable $j$, than for some $i$ the following logical statement will be true $(\average{\overline{Z^{(N),j}_{i-1}}}=-1) \wedge (\average{\overline{Z^{(N),j+1}_{i}}}=1)$. Therefore we can use interactions of the form $(1+\overline{Z^{(N),j}_{i-1}})\,(1-\overline{Z^{(N),j+1}_{i}})$ to enforce the constraint, $(\average{\overline{Z^{(N),j}_{i-1}}}=1) \vee (\average{\overline{Z^{(N),j+1}_{i}}}=-1) \forall i$, summing over $i$ (and neglecting an irrelevant constant offset) we obtain
\begin{equation}
H^{(j,j+1)}_{>}=\sum_{i=0}^{N-1}-\overline{Z^{(N),j}_{i-1}}\,\overline{Z^{(N),j+1}_i}-\overline{Z^{(N),j+1}_i}+\overline{Z^{(N),j}_{i-1}}
\end{equation}
which yields the minimum possible energy if the value of  variable $j$ is less than that of $j+1$, and a positive energy otherwise \footnote{This expression can even be made more efficient by omitting domain wall sites corresponding to `impossible' values, for instance, there is no legal configuration where the $j$th variable can take a value less than $j$}. The $k$-hot condition can therefore be enforced by adding $\lambda' \sum_{j=0}^{j=k-2} H^{(j,j+1)}_{>}$ to the domain wall Hamiltonians which implement $k$ variables.

Constructing domain wall analogues of $k$-hot problems seems counter-productive, since it requires more logical qubits than the analogous $k$-hot encoding. However, in these $k$-hot analogies, the parts of the domain walls which encode different discrete values can still be spatially separated, for some problem, the domain wall analogue may still be more efficient after embedding. While a potentially fruitful endeavour, examining the efficiency trade-offs between $k$-hot and domain wall $k$-hot analogues for encoding real problems is beyond the scope of this paper.

\section{Discussion and conclusions\label{sec:discuss}}

In this manuscript, I have discussed a method of encoding discrete variables into Ising model qubits based on domain walls in one dimensional spin chains which is an alternative to the traditional one hot method. I have further demonstrated how arbitrary (classical) two variable interactions can be encoded, and that these interactions only require two body Ising terms. Furthermore, as was demonstrated in \cite{Hadfield17a} for one hot, two body mixer terms which preserve the logically valid space are possible, which may be highly relevant in QAOA implementations \cite{Hadfield17a}. Finally, I have numerically examined the possibility of embedding problems encoded using the domain wall methods in two graphs which are relevant to quantum annealing, the chimera and Pegasus graphs. For every problem type I have examined, I found that the domain wall encoding can be embedded more efficiently than the one hot encoding. The level of improvement is strongly dependent on problem type, but in some cases can be comparable to the gains made from having a Pegasus versus chimera hardware graph. Specifically, for the synthetic scheduling problems examined here,embedded into the Pegasus graph, the gains made in terms of the ratio of physical to logical bits for domain wall versus one hot is comparable to the gains from embedding into Pegasus versus chimera. 

It is likely that the large gains are due the fact that the domain wall encoding inherently allows variables to be more `spread out' and therefore take advantage of the natural structure of the problem for embedding in a way which is not possible for one hot. One example of such structure is the fact that events occurring at very different times in a scheduling problem are unlikely to interact. For two such interacting variables in one hot, the binary variables representing the events happening even at very different times can have a maximum edge distance of two (two edges must be traversed to get between them), whereas for a domain wall encoding, the edge distance is in principle unbounded. It is worth emphasising that this study was performed using general purpose heuristic problem embedding software, and therefore it is likely that even better results could be obtained using specialized software which is specifically designed to take advantage of known structural features within specific problems. Furthermore, the nominal hardware graphs studies here are also general purpose, it is likely that embedding overhead could be reduced or even eliminated on application specific hardware, for instance ASICs designed for problems with a particular structure. 

It would further be interesting to examine the most efficient strategies to decompose problems which do not fit onto the hardware graph of a quantum annealer\cite{Bian14a,Bian16a,qbsolve}. The importance of such strategies has been highlighted in more general cases \cite{Okada19a}.

It is finally worth briefly noting that the natural structure present in the domain wall encoding means that it can be used to design discrete or mixed binary/integer optimization problems which can be mapped to hardware with no embedding overhead. This could provide an important tool for scientific studies on real quantum annealers, as embedding may complicate the interpretation of experimental results. In particular, in an upcoming work \cite{chancellor19a} I will examine the ability of real quantum annealers to find solutions to mixed binary/integer problems which are not only highly optimal but also robust, and examine how reverse annealing \cite{reverse_anneling_whitepaper,chancellor17b} can be used as a tool to trade off between optimality and robustness.
 
\section*{Acknowledgments}

This work was supported by BP plc, by EPSRC fellowship EP/S00114X/1, and ESPRC Hub EP/M013243/1. Figures were drawn using the Tikzit tikz editor \cite{tikzit}. Numerical calculations were performed in Python 3.5 \cite{van2003python} using jupyter notbeooks \cite{jupyter,perez2007ipython} and used the numpy \cite{oliphant2006guide},  minorminer \cite{minorminer}, D-Wave networkx \cite{D-Wave_nx} modules. Plotting was performed using the maplotlib module \cite{hunter2007matplotlib}. The author thanks Viv Kendon for a critical reading of the paper, Jie Chen for spotting several typos in formulae posted in ar$\chi$iv versions, and Adam Callison for pointing out that quadratic interactions can be implemented efficiently for a binary encoding.

\section*{Appendix 1: Explantion of performance metrics given in table \ref{tab:comparison}}
We use several performance metrics to compare binary, one-hot, and domain wall encodings in table \ref{tab:comparison}, in this appendix. We also explain what makes the cases labelled as `complicated' complicated, and why no simple answer can be given. 

\subsection*{\# qubits}
This is the number of qubits which each encoding requires to logically represent the problem. In the case of binary encoding, this excludes any auxiliary qubits required to implement constraints, see `\# couplers for encoding' for a discussion of where such qubits may come about.

\subsection*{\# couplers for encoding}
This is the number of two body couplers required to restrict the qubits to the \emph{logically valid} subspace, in other words, the subspace of qubit configurations which correspond to logically valid values of the discrete variable being encoded. In the case of domain wall and one hot, this is straightforward, since the number of couplers for each is well defined. In the case of binary encoding, the complexity depends strongly on whether or not the discrete variable being encoded involves a binary (i.e.~$m=2^n,\,m \in \mathbb{Z}$) number of possibilities or not. If it does, than the logically valid space is exactly the space of all possible bitstrings expressible in $n$ qubits, and therefore no couplers are required to constrain the space. On the other hand if the number of possibilities is not a binary number, than constraints need to be added to prevent certain configurations, in general, these constraints will involve more than two body terms and would therefore also require auxiliary qubits to implement effective multi-body constraints. A full analysis of all of the ways this can be done is beyond the scope of this paper, so this case has been simply marked as `complicated'.

\subsection*{intra-variable connectivity}
This is the structure of the connectivity required within the variable to restrict to the logically valid subspace. For one hot this requires a fully connected construction, and for domain wall it requires linear connectivity. As with the number of couplers, the connectivity depends whether or not the discrete variable involves a binary number of possibilities, if it does, than no connections at all are needed to restrict to the logically valid subspace. On the other hand if there is not a binary number, than some configurations need to be excluded, which will generally require high order coupling, the structures necessary to do this are complicated and beyond the scope of this paper.

\subsection*{maximum order needed to penalize single values}
This is the highest order of coupling (i.e.~how many qubits need to be coupled together in a single effective interaction) to penalize an arbitrary single logical value. For both one hot and domain wall, this is one, since single body terms which either act on a single qubit or bracket a domain wall respectively can achieve this. For a binary encoding, one would in general need interactions which interact every qubit used to encode the variable with every other qubit used to encode the variable. Note that some specific kinds of penalties in the binary encoding may require much lower order, for instance, a penalty which scales linearly with the value of a binary number only requires single body terms.

\subsection*{maximum order needed for two variable interactions}
This is the highest order of coupling (i.e.~how many qubits need to be coupled together in a single effective interaction) to implement an \emph{arbitrary} two variable interaction. For both one hot and domain wall encodings this is two. For the binary encoding, this could require a coupling which involved every qubit in both variables. Specific interactions can require lower order though, for instance a quadratic interaction can be encoded using only one and two body interactions.

\subsection*{maximum $d_e$ between qubits in interacting variables}
This is the maximum edge distance between qubits used to encode two different interacting variables. The edge distance is the minimum number of edges which must be traversed to get between two qubits in the interaction graph. Roughly speaking this measures how `spread out' a variable encoding can be. As discussed previously in this appendix, interactions between binary variables can be complex to encode and may require auxiliary qubits if native high order interactions are not available, therefore the edge distance in the binary case is complicated and case dependant.

\section*{Appendix 2: Hamiltonians for examples}

In this appendix, I give explicit Hamiltonians for all of the example problem types. To simplify the expressions and to make it so the same Hamiltonians can be used to express both types of domain wall encodings as well as one hot, it is useful to provide some definitions. First of all, I define the variable cores which are the constraints necessary to define the variable for domain walls are defined by Eq.~\ref{eq:dw_Ham_Zbar}
\begin{equation}
H^{(N),k}_{\mathrm{core}}=-\lambda \sum_{i=-1}^{N-1} \overline{Z^{(N),k}_i}\, \overline{Z^{(N),k}_{i+1}},
\end{equation}
where the notation has been modified to add the variable index $k$, and to clarify the role of the Hamiltonian. Similarly, following from Eq.~\ref{eq:one_hot}, the core Hamiltonian in the one hot equation can be defined as
\begin{equation}
H^{(N),k}_{\mathrm{core}}=\lambda \left(\sum_{i<j}Z^k_iZ^k_j-(m-1)\sum_iZ^k_i\right).
\end{equation}
In addition to the core, I need to define penalties on different variable values, for the domain wall encoding, this comes from Eq.~\ref{eq:delta_def}
\begin{equation}
\delta^{(N),k}_i=\frac{1}{2} (\overline{Z^{(N)}_{i}}-\overline{Z^{(N)}_{i-1}})
\end{equation}
where the bar has been dropped for notational convenience. Finally, for one hot, the definition is very simple, 
\begin{equation}
\delta^{(N),k}_i=\frac{1}{2} Z^k_{i+1}.
\end{equation}
Equipped with these definitions, I now define the Hamiltonians for the three examples

\subsection{Unstructured Interactions}

In this case I consider unstructured interactions between two variables, although the definition can be easily extended to more. In this case, let the interactions between variable $1$ of size $N$, and variable $2$ of size $M$ be defined by the $N\times M$ matrix $A$. The Hamiltonian is therefore
\begin{equation}
H_{\mathrm{unstruct}}=H^{(N),1}_{\mathrm{core}}+H^{(M),2}_{\mathrm{core}}+\sum_{i=1}^N\sum_{j=1}^M A_{ij}\delta^{(N),1}_{i-1}\delta^{(M),2}_{j-1}.
\end{equation}

\subsection{Graph colouring}

In this case, we consider colouring a graph with $K$ vertexes and edges defined by the $K\times K$ strictly upper triangular matrix $e$ where $1$ denotes an edge and $0$ represents no edge, using $N$ colours. Each variable will be of size $N$ and will have constraints which prevent vertexes which are coloured the same to share edges.  The Hamiltonian therefore takes the form
\begin{equation}
H_{\mathrm{color}}=\sum_{l=1}^K H^{(N),l}_{\mathrm{core}}+\sum_{i=1}^{K}\sum_{j=i+1}^{K}e_{ij}\sum_{k=0}^{N-1}\delta^{(N),i}_{k}\delta^{(M),j}_{k}.
\end{equation}

\subsection{Scheduling}

In this case I consider a scheduling problem which is defined to involve $m$ events. A single event, event $k$, must occur between $t_{k,\mathrm{min}}$ and $t_{k,\mathrm{max}}$, and furthermore has a duration $T_k$. For mathematical convenience, I define $\mathrm{dur}(k)=t_{k,\mathrm{max}}-t_{k,\mathrm{min}}$ Some events conflict, meaning that they cannot occur simultaneously while others do not. I define whether or not events conflict using the strictly upper triangular matrix $C$, where $1$ represents a conflict and $0$ represents no conflict. The Hamiltonian is defined as
\begin{align}
H_{\mathrm{sched}}=\sum_{k=1}^m H^{(\mathrm{dur}(k)),k}_{\mathrm{core}}+ \nonumber \\
\sum_{i=1}^{m}\sum_{j=i+1}^{m}C_{ij}\sum_{l=1}^{\mathrm{dur}(i)}\sum_{q=1}^{\mathrm{dur}(j)}R^{(i,j)}_{lq}\delta^{(\mathrm{dur}(i)),i}_{l-1}\delta^{(\mathrm{dur}(j)),j}_{q-1}.
\end{align}
The binary matrix $R^{(i,j)}_{lq}$ defines whether or not two events temporally overlap given their duration, in other words, 
\begin{widetext}
\begin{equation}
R^{(i,j)}_{lq}=\begin{cases} 1 & t_{i,\mathrm{min}}+l-1=t_{j,\mathrm{min}}+q-1 \\ 1 & t_{j,\mathrm{min}}+q-1<t_{i,\mathrm{min}}+l-1<t_{j,\mathrm{min}}+q-1+T_j \\ 1 & t_{i,\mathrm{min}}+l-1<t_{j,\mathrm{min}}+1-1<t_{i,\mathrm{min}}+q-1+T_i \\ 0 & \mathrm{otherwise}\end{cases}.
\end{equation}
\end{widetext}

\bibliography{bibLibrary}  

\end{document}